\documentclass[12pt]{article}
\pdfoutput=1

\usepackage[utf8]{inputenc}
\usepackage[left=2.55cm, right=2.55cm, top=2.55cm, bottom=2.55cm]{geometry}
\usepackage{amsmath,amssymb,amsbsy}
\usepackage{slashed}
\usepackage{xcolor}
\usepackage{graphicx}
\usepackage{url}
\usepackage{cancel}
\usepackage{cite}
\usepackage[colorlinks=true,allcolors=blue,pdfborder={0 0 0},linktocpage=false,pdfencoding=auto]{hyperref}
\usepackage{tabularx,booktabs}
\usepackage{multicol}
\usepackage{feynmp}
\usepackage{units}
\usepackage{xspace}
\usepackage[labelfont=bf]{caption}
\usepackage[section]{placeins}
\usepackage{subcaption}
\usepackage{soul} 
\usepackage{bbold}
\usepackage{parskip}
\usepackage{float}
\usepackage{tabulary}

\definecolor{darkred}{rgb}{0.6,0,0}
\definecolor{darkpurple}{rgb}{0.5,0,0.5}

\newcommand{\code}[1]{\texttt{#1}}
\newcommand{\beqn}{\begin{eqnarray}}
\newcommand{\eeqn}{\end{eqnarray}}

\def\amu{$\Delta{a^{\rm FB}_\mu}$}

\begin{document}

\author{Amin Aboubrahim$^a$\footnote{\href{mailto:aabouibr@uni-muenster.de}{aabouibr@uni-muenster.de}}~, Pran Nath$^b$\footnote{\href{mailto:p.nath@northeastern.edu}{p.nath@northeastern.edu}}~ and Raza M. Syed$^c$\footnote{\href{mailto:rsyed@aus.edu}{rsyed@aus.edu}} \\~\\
$^{a}$\textit{\normalsize Institut f\"ur Theoretische Physik, Westf\"alische Wilhelms-Universit\"at M\"unster,} \\
\textit{\normalsize Wilhelm-Klemm-Stra{\ss}e 9, 48149 M\"unster, Germany} \\
$^{b}$\textit{\normalsize Department of Physics, Northeastern University, Boston, MA 02115-5000, USA} \\
$^{c}$\textit{\normalsize Department of Physics, American University of Sharjah, P.O. Box 26666, Sharjah, UAE\footnote{Permanent address}}
\\}

\title{\vspace{-2cm}\begin{flushright}
{\small MS-TP-21-12}
\end{flushright}
\vspace{2cm}
\Large \bf
Yukawa coupling unification in an $\mathsf{SO(10)}$ model consistent with Fermilab $(g-2)_{\mu}$ result
 \vspace{0.5cm}}

\date{}
\maketitle

\vspace{1cm}

\begin{abstract}
We investigate the Yukawa coupling unification for the third generation in a class of $\mathsf{SO(10)}$
unified models which are consistent with the 4.2 $\sigma$ deviation from the standard model of the muon $g-2$
seen by the Fermilab experiment E989.  A recent analysis in supergravity grand unified models shows that such an effect can arise from supersymmetric loops correction. Using a neural network, we further
analyze regions of the parameter space where Yukawa coupling unification consistent with the Fermilab
result can appear. In the analysis we take into account the contributions to Yukawas from the
cubic and the quartic interactions. We test the model at the high luminosity
and high energy LHC and estimate the integrated luminosities needed to discover sparticles
predicted by the model.

\end{abstract}

\numberwithin{equation}{section}

\newpage

{  \hrule height 0.4mm \hypersetup{colorlinks=black,linktocpage=true} \tableofcontents
\vspace{0.5cm}
 \hrule height 0.4mm}

\section{Introduction}\label{sec1}
\noindent

Recently the Fermilab E989 experiment~\cite{Abi:2021gix} has measured $a_\mu=(g-2)_\mu/2$ with
significantly greater accuracy than the previous Brookhaven experiment~\cite{Bennett:2006fi,Tanabashi:2018oca}.
Thus the combined Fermilab experimental data and Brookhaven experimental data gives
 \begin{equation}
 a^{\rm exp}_\mu = 116 592 061 (41)  \times10^{-11} \,,
  \label{Brook}
\end{equation}
which is to be compared with the Standard Model (SM) prediction~\cite{Aoyama:2020ynm}
 \begin{equation}
 a^{\rm SM}_\mu = 116 591 810 (43)  \times10^{-11}.
  \label{SM}
\end{equation}
The combined Fermilab and Brookhaven result shows an excess over the SM result
by an amount $\Delta a^{\rm FB}_\mu$ which is
  \begin{equation}
\Delta a^{\rm FB}_\mu = a_{\mu}^{\rm exp} - a_{\mu}^{\rm SM} = 251 (59)  \times10^{-11}.
  \label{diff}
\end{equation}
Eq.~(\ref{diff}) records a $4.2\sigma$ deviation from the SM compared to $3.7 \sigma$
for the Brookhaven result. Thus the Fermilab experiment further strengthens the Brookhaven result
on the possible existence of new physics beyond the Standard Model.
Subsequent to the Fermilab result,  artificial neural network analysis was used to explore the
parameter space of supergravity (SUGRA) unified models. It was seen that regions of the parameter space
where supersymmetric loops can give the desired correction consistent with the Fermilab results
are those where gluino-driven radiative breaking of the electroweak symmetry occurs~\cite{Aboubrahim:2021rwz}, a region referred to as $\tilde g$SUGRA~\cite{Akula:2013ioa,Aboubrahim:2019vjl,Aboubrahim:2020dqw}.
 Using a neutral network we investigate this region further to explore
the region where Yukawa unification in an $\mathsf{SO(10)}$ model~\cite{Aboubrahim:2020dqw}
can occur consistent with the Fermilab result.

The outline of rest of the paper is as follows: In  section~\ref{sec2} details of the $\mathsf{SO(10)}$
model are discussed. In section~\ref{sec3} an analysis of the parameter space of
SUGRA $\mathsf{SO(10)}$ model which gives Yukawa coupling unification consistent with the Fermilab $g-2$ result
is given. Here the light and the heavy sparticle spectrum is also computed.
In section~\ref{sec4}, simulations  for the observation 
of the sparticles predicted by the model  at HL-LHC and HE-LHC are given.
Conclusions are given in section~\ref{sec5}. Some further details of the model are given
in the Appendix.

\section{The model \label{sec2}}
  The general class of $\mathsf{SO(10)}$ models we consider are those of~\cite{Babu:2006nf,Babu:2011tw}
  with~\cite{Aboubrahim:2020dqw} being one of them which are similar in spirit to the missing partner
  $\mathsf{SU(5)}$ models~\cite{Masiero:1982fe,Grinstein:1982um}.
   These models  involve large
  Higgs representations such as $\mathsf{126+\overline{126}}$, $\mathsf{210}$, $\mathsf{120}$
  for Yukawa  couplings. Large Higgs representations have been used in several  early  works~\cite{Clark:1982ai,Aulakh:1982sw,Babu:1992ia}
   and also more recently, e.g.,~\cite{Nath:2001uw,Nath:2001yj,Nath:2003rc,Aulakh:2003kg,Bajc:2004xe,Aulakh:2004hm,Aulakh:2008sn}
and the references therein (for a review of $\mathsf{SO(10)}$ models, see Ref.~\cite{Nath:2006ut}).
  In the model we
 consider~\cite{Babu:2011tw,Aboubrahim:2020dqw},  the missing
 partner mechanism comes about as follows: The Higgs sector consists of the fields
 $\mathsf{126+\overline{126}}$, $\mathsf{210}$, $2\times \mathsf{10 +120}$ set of representations.
 The fields $\mathsf{126+\overline{126}}$, $\mathsf{210}$ are heavy which break the
 GUT symmetry down to the SM gauge group symmetry, while the $2\times \mathsf{10 +120}$ Higgs
 fields  are light.  The heavy fields contain 3 pairs of heavy Higgs doublets while the light fields have
  four pairs of light Higgs doublets. When the light and heavy fields mix, three pairs of the
  light Higgs doublets become heavy while one combination of the light doublets remains light
  and is identified as the Higgs field of the MSSM. We give below further details of the model used
  in this analysis.

The superpotential of the $\mathsf{SO(10)}$ model is given by~\cite{Aboubrahim:2020dqw}
\begin{align}
W=W_{\rm GUT} + W_{\rm DT} + W_{\rm Yuk},
\end{align}
where
\begin{align}
\label{gut}
W_{\textsc{gut}}=&~ M^{126}\Delta_{\mu\nu\rho\sigma\lambda}\overline{\Delta}_{\mu\nu\rho\sigma\lambda}+M^{210}\Phi_{\mu\nu\rho\sigma}\Phi_{\mu\nu\rho\sigma}
+\eta\Phi_{\mu\nu\rho\sigma}\Delta_{\mu\nu\lambda\tau\xi}\overline{\Delta}_{\rho\sigma\lambda\tau\xi} \nonumber \\
&+\lambda \Phi_{\mu\nu\rho\sigma} \Phi_{\rho\sigma\lambda\tau} \Phi_{\lambda\tau\mu\nu}\,,\\
W_{\textsc{dt}}=~& a~{^1}\Omega_{\mu}\overline{\Delta}_{\mu\nu\rho\sigma\lambda}\Phi_{\nu\rho\sigma\lambda}+\sum_{r=1}^2 b_r~{^r}\Omega_{\mu}{\Delta}_{\mu\nu\rho\sigma\lambda}\Phi_{\nu\rho\sigma\lambda}+c~ \Sigma_{\mu\nu\rho}\Delta_{\nu\rho\sigma\lambda\tau}\Phi_{\mu\sigma\lambda\tau}\nonumber\\
&+
\overline{c}~ \Sigma_{\mu\nu\rho}\overline{\Delta}_{\nu\rho\sigma\lambda\tau}\Phi_{\mu\sigma\lambda\tau}\,.
\label{dtsuper}
\end{align}
The notation used above is as follows:
$\Delta_{\mu\nu\rho\sigma\lambda}$ and  $\overline{\Delta}_{\mu\nu\rho\sigma\lambda}$
are fields for the  $\mathsf{126}$ and  $\mathsf{\overline{126}}$ representations,
$\Phi_{\mu\nu\rho\sigma}$ is the field for the $\mathsf{210}$ representation and
${^r}\Omega_{\mu} (r=1,2)$ are the fields for the two $\mathsf{10}$ of Higgs representations
and $\Sigma_{\mu\nu\rho}$ is the field for the $\mathsf{120}$-plet representation.
In the above
 $W_{\rm GUT}$ breaks the $\mathsf{SO(10)}$ GUT symmetry down to the standard model gauge group
$\mathsf{SU(3)_C\times SU(2)_L\times U(1)_Y}$  by VEV formations of  $\mathcal V_{1_{126}}$ and $\mathcal V_{1_{\overline{126}}}$ and the VEVs of
$\mathcal V_{1_{210}}$, $\mathcal V_{24_{210}}$, $\mathcal V_{75_{210}}$.
The equations that determine these VEVs are derived in \cite{Aboubrahim:2020dqw}.
Thus the $\mathsf{126+\overline{126}}$-plet
VEVs  $\mathcal V_{1_{126}}$ and $\mathcal V_{1_{\overline{126}}}$ break the $\mathsf{SO(10)}$
symmetry down to $\mathsf{SU(5)\times U(1)}$ and the  $\mathsf{210}$-plet VEVs
$\mathcal V_{1_{210}}$, $\mathcal V_{24_{210}}$, $\mathcal V_{75_{210}}$ further break the gauge
symmetry down to $\mathsf{SU(3)_C\times SU(2)_L\times U(1)_Y}$.
The notation for the VEVs is explicit. Thus, for example,  $\mathcal V_{1_{126}}$ stands for the VEV of the
$\mathsf{SU(5)}$ singlet in the $\mathsf{SU(5)\times U(1)}$ decomposition of $\mathsf{126}$ and
$\mathcal V_{24_{210}}$ stands for the VEV of the the $\mathsf{24}$-plet of $\mathsf{SU(5)}$ field in the
$\mathsf{SU(5)\times U(1)}$ decomposition of $\mathsf{210}$.
The doublet-triplet splitting is generated by $W_{\rm DT}$ which contains $2\times \mathsf{10 +120}$-plets
of light fields. Thus the heavy fields
$\mathsf{126+\overline{126}}$-plet and $\mathsf{210}$-plet contain three heavy $\mathsf{SU(2)}$ Higgs
doublet pairs while the light fields $2\times \mathsf{10 +120}$-plets contain four light Higgs doublet pairs.
After mixing of the light and heavy fields, three light Higgs doublets become heavy leaving one
pair massless which we identify as the standard model Higgs doublet.

The Yukawa couplings arise from cubic and quartic interactions. They are given by
\begin{align}
W_{\rm Yuk}= W_3 + W_4,
\label{wyuk}
\end{align}
where
\begin{equation}
\label{w3}
W_3=\sum_{r=1}^2f^{10_{r}}~\langle\Psi_{(+)}^*|B \Gamma_{\mu}|\Psi_{(+)}\rangle~ {^r}{\Omega_{\mu}}\,.
\end{equation}
Here $B$ and $\Gamma$'s are the $\mathsf{SO(10)}$ charge conjugation and gamma matrices~\cite{Nath:2001uw} and $W_4$ are the higher dimensional interactions
 discussed below.
Yukawa couplings arising from Eq.~(\ref{w3}) are given by
\begin{eqnarray}
\mathcal{L}_{\textnormal{Yuk}}= +h^0_{\tau}~\epsilon^{ab}{\mathbf{H_d}}_a\mathbf{{L}}_{b}{\mathbf E}^{\mathtt c}
-h^0_{b}~{\mathbf{H_d}}_a\mathbf{{Q}}^{a\alpha}{\mathbf D}_{\alpha}^{\mathtt c}-h^0_{t}~\epsilon_{ab}{\mathbf{H_u}}^a\mathbf{{Q}}^{b\alpha}{\mathbf U}_{\alpha}^{\mathtt c} + \textnormal{h.c.},
\end{eqnarray}
 where
\begin{eqnarray}\label{q&l masses from cubic coupling}
h^0_{\tau}&=i2\sqrt{2}\sum_{r=1}^2f^{10_{r}}V_{d_{r1}},~~
h^0_{b}=-i2\sqrt{2}\sum_{r=1}^2f^{10_{r}}V_{d_{r1}}, ~~
h^0_{t}=-i2\sqrt{2}\sum_{r=1}^2f^{10_{r}}U_{d_{r1}},
\end{eqnarray}
where $U_{d_{r1}}$ and $V_{d_{r1}}$ are defined by  Eq.~(\ref{UV})
and  evaluated numerically in Tables~\ref{tab2} and~\ref{tab3}.  In addition to Yukawa couplings arising
from $W_3$, contributions arise from higher dimensional operators in $W_4$ where
\begin{align}
W_4=W^{(1)}_4+ W^{(2)}_4+ W^{(3)}_4,
\end{align}
and where
\begin{eqnarray}
W^{(1)}_4&=&-\frac{f^{(1)}}{5!M_c}b_r\langle\Psi_{(+)}^*|B \Gamma_{[\lambda}\Gamma_\mu\Gamma_{\nu}\Gamma_{\rho}\Gamma_{\sigma]}|\Psi_{(+)}\rangle~
\left[{^r}\Omega_{\lambda}\Phi_{\mu\nu\rho\sigma}-{^r}\Omega_{\mu}\Phi_{\lambda\nu\rho\sigma}+{^r}\Omega_{\nu}\Phi_{\lambda\mu\rho\sigma}\right.\nonumber\\
&&\left.~~~~~~~~~~~~~~~~~~~~~~~~~~~~~~~~~~~~~~~~~~~~~~~~~~~-{^r}\Omega_{\rho}\Phi_{\lambda\mu\nu\sigma}+{^r}\Omega_{\sigma}\Phi_{\lambda\mu\nu\rho}\right]\,,
\label{w41}
\\
W_4^{(2)}&=&-\frac{f^{(2)}}{5!M_{c}}\langle\Psi_{(+)}^*|B \Gamma_{[\lambda}\Gamma_\mu\Gamma_{\nu}\Gamma_{\rho}\Gamma_{\sigma]}|\Psi_{(+)}\rangle~
\left[\Sigma_{\lambda\alpha\beta}\Phi_{\gamma\rho\sigma\lambda}-\Sigma_{\lambda\alpha\gamma}\Phi_{\beta\rho\sigma\lambda}
+\Sigma_{\lambda\alpha\rho}\Phi_{\beta\gamma\sigma\lambda}\right.\nonumber\\
&&\left.~~~~~~~~~~~~~~~~~~~~~~~~~~~~~~~~~~~~~~~~~~~~~~~~-\Sigma_{\lambda\alpha\sigma}\Phi_{\beta\gamma\rho\lambda}
-\Sigma_{\lambda\gamma\beta}\Phi_{\alpha\rho\sigma\lambda}
+\Sigma_{\lambda\rho\beta}\Phi_{\alpha\gamma\sigma\lambda}\right.\nonumber\\
&&\left.~~~~~~~~~~~~~~~~~~~~~~~~~~~~~~~~~~~~~~~~~~~~~~~~-\Sigma_{\lambda\sigma\beta}\Phi_{\alpha\gamma\rho\lambda}
-\Sigma_{\lambda\gamma\rho}\Phi_{\beta\alpha\sigma\lambda}
+\Sigma_{\lambda\gamma\sigma}\Phi_{\beta\alpha\rho\lambda}\right.\nonumber\\
&&\left.~~~~~~~~~~~~~~~~~~~~~~~~~~~~~~~~~~~~~~~~~~~~~~~~-\Sigma_{\lambda\rho\sigma}\Phi_{\beta\alpha\gamma\lambda}\right]\,, \label{w42}
\\
W_4^{(3)}&=&\frac{f^{(3)}}{M_{c}}\langle\Psi_{(+)}^*|B \Gamma_{\mu}|\Psi_{(+)}\rangle\Sigma_{\rho\sigma\lambda}\Phi_{\rho\sigma\lambda\mu}\,.
\label{w43}
\end{eqnarray}

Thus $W_4$ gives additional contributions to the Yukawa couplings for the third generation
which we denote by $\delta h_t,~ \delta h_b,~ \delta h_\tau$ which are evaluated in the Appendix.
The total Yukawa couplings arising from Eq.~(\ref{wyuk}) is then given by
\begin{align}
h_t=h^0_t +\delta  h_t,~~h_b=h^0_b +\delta  h_b,~~
h_\tau=h^0_\tau +\delta  h_\tau\,,
\label{yuksum}
\end{align}
where
 $h_b,~ h_t,~h_\tau$ act as boundary conditions on Yukawas of $b,t,\tau$ which are
 evolved down to the electroweak scale $Q$ where they are related to
$b,t,\tau$ masses so that
\beqn\label{yukawa-1}
m_{t}(Q)=\frac{h_{t}(Q) v \sin\beta}{\sqrt 2}, ~~m_{b}(Q)=\frac{h_{b}(Q) v \cos\beta}{\sqrt 2},
~~m_{\tau}(Q)=\frac{h_{\tau}(Q) v \cos\beta}{\sqrt 2}.
\eeqn
Here we used the relations $\langle H_d\rangle=\frac{v}{\sqrt 2} \cos\beta$ and $\langle H_u\rangle=\frac{v}{\sqrt 2}\sin\beta$, and where $v=246$ GeV.

As noted above there are seven Higgs doublet pairs three of which are heavy and four are light,
and after the mixing of the light and heavy fields three pairs of light Higgs doublets become
heavy and one pair remains light. To extract the light Higgs doublets we need to diagonalize the
$7\times 7$ Higgs doublet mass matrix given  in~\cite{Aboubrahim:2020dqw}.
The Higgs doublet mass matrix is not symmetric and is diagonalized by two unitary matrices $U_d$ and $V_d$.
Thus the down Higgs  and the up Higgs doublet mass  matrices are diagonalized by the transformation
\begin{align}
{\cal H}_d= V_d {\cal H}_d', ~~{\cal H}_u = U_d {\cal H}_u'\,,
\label{UV}
\end{align}
where
\begin{eqnarray}\label{doublet mass eigenstates}
 {\cal H}_d^T&=&  ({}^{(\overline{5}_{10_1})}\!{\mathsf D}_{a},  {}^{(\overline{5}_{10_2})}\!{\mathsf D}_{a}, {}^{(\overline{5}_{120})}\!{\mathsf D}_{a}, {}^{(\overline{5}_{{126}})}\!{\mathsf D}_{a},{}^{(\overline{5}_{{210}})}\!{\mathsf D}_{a}, {}^{(\overline{45}_{120})}\!{\mathsf D}_{a}, {}^{(\overline{45}_{\overline{126}})}\!{\mathsf D}_{a}),\\\
  {\cal H}_d^{'T}&=&
 ( {\mathbf{H_d}}_a, {}^{2}\!{\mathsf D}_{a}^{\prime},
 {}^{3}\!{\mathsf D}_{a}^{\prime}, {}^{4}\!{\mathsf D}_{a}^{\prime}, {}^{5}\!{\mathsf D}_{a}^{\prime}, {}^{6}\!{\mathsf D}_{a}^{\prime}, {}^{7}\!{\mathsf D}_{a}^{\prime}),\\
{\cal H}_u^T&=&  ({}^{(\overline{5}_{10_1})}\!{\mathsf D}^{a},  {}^{(\overline{5}_{10_2})}\!{\mathsf D}^{a}, {}^{(\overline{5}_{120})}\!{\mathsf D}^{a}, {}^{(\overline{5}_{{126}})}\!{\mathsf D}^{a},{}^{(\overline{5}_{{210}})}\!{\mathsf D}^{a}, {}^{(\overline{45}_{120})}\!{\mathsf D}^{a}, {}^{(\overline{45}_{\overline{126}})}\!{\mathsf D}^{a}),\\
   {\cal H}_u^{'T}&=&
 ( {\mathbf{H_u}}^a, {}^{2}\!{\mathsf D}^{a\prime},
 {}^{3}\!{\mathsf D}^{a\prime}, {}^{4}\!{\mathsf D}^{a\prime}, {}^{5}\!{\mathsf D}^{a\prime}, {}^{6}\!{\mathsf D}^{a\prime}, {}^{7}\!{\mathsf D}^{a\prime}).
  \end{eqnarray}
  In the above the notation is as follows: ${}^{(\overline{5}_{10_1})}$ stands for the down Higgs doublet
  in the $\mathsf{SU(5)}-\overline{\mathsf{5}}$-plet  in the $\mathsf{10}_1$ which is one of the two $\mathsf{10}$-plets of  light Higgs
  of $\mathsf{SO(10)}$.
  Further,
     $\mathsf D$'s and ${\mathsf D}^{\prime}$'s represent the normalized kinetic energy basis and normalized kinetic and mass eigenbasis, respectively of the Higgs doublet mass matrix. The pair of doublets $({\mathbf{H_d}}_a,{\mathbf{H_u}}^a)$ are identified to be light and are the
 normalized   electroweak Higgs doublets of the minimal supersymmetric standard model (MSSM).
 The matrix elements of $U_d$ and $V_d$ relevant in our analysis below are those elements that connect
 the light doublets, i.e., $U_{d_{11}}, U_{d_{21}}, \cdots. U_{d_{71}}$, and the elements
 $V_{d_{11}}, V_{d_{21}}, \cdots, V_{d_{71}}$. Other
matrix elements of $U_d$ and $V_d$  do not contribute in the low energy theory.
As noted above
the explicit form of the $7\times 7$ Higgs doublet mass matrix is given in~\cite{Aboubrahim:2020dqw}.
The $U$ and the $V$ matrices are obtained by diagonalization of this matrix.
Numerical values of the non-zero matrix elements of  $U_d$ and $V_d$ relevant in the analysis
are displayed in Tables \ref{tab2} and \ref{tab3} for benchmarks of Table \ref{tab1}.

\section{$\mathsf{SO(10)}$ SUGRA model with Yukawa unification consistent with Fermilab $(g-2)_{\mu}$\label{sec3}}

Since the muon $g-2$ is one of the most accurately determined quantities in physics
even a small deviation from the standard model prediction would be a significant indicator
of new physics. For example, it is known that supersymmetric loop corrections
could be of the same size as the electroweak corrections in the SM~\cite{Kosower:1983yw,Yuan:1984ww,Lopez:1993vi,Chattopadhyay:1995ae,Moroi:1995yh,Carena:1996qa}.
Indeed the Brookhaven result in 2001~\cite{Bennett:2006fi} resulted in several works
pointing out the impact on physics expected at colliders and elsewhere~\cite{Czarnecki:2001pv,Chattopadhyay:2001vx,Everett:2001tq,Feng:2001tr,Baltz:2001ts,Sabatta:2019nfg,vonBuddenbrock:2019ajh,Chen:2021rnl}.
Thus the experiment became one of the important constraints on the parameter space of
SUSY models. The discovery of the Higgs boson at 125 GeV further constrained the parameter
space implying that the size of weak  SUSY scale could be large lying in the TeV
region~\cite{Akula:2011aa,higgs7tev1}. Since the Fermilab result has indicated more
strongly than the Brookhaven experiment for the existence of new physics, it is interesting
to ask how the $b-t-\tau$ unification is affected~\cite{Abi:2021gix}.  The early work
of~\cite{Ananthanarayan:1992cd} pointed out that such a unification could occur
in $\mathsf{SO(10)}$ with appropriate choice of soft parameters. Such a unification has important
effects on other phenomena such as dark matter (DM)~\cite{Chattopadhyay:2001va}.
Thus it is of interest to
ask if $b-t-\tau$ unification can come about  consistent with Fermilab data.
We investigate this question using a neural network which is found to be useful in the
analysis of large parameter spaces~\cite{Hollingsworth:2021sii,Balazs:2021uhg}).

The analysis is done within the framework of supergravity grand unified models~\cite{sugrauni}
using non-univeralities of gaugino masses~\cite{Nath:1983iz,Ellis:1985jn,nonuni2,Feldman:2009zc,Belyaev:2018vkl}.
The scan of the SUGRA parameter space is performed using an artificial neutral network (ANN) implemented in \code{xBIT}~\cite{Staub:2019xhl}. The ANN has three layers with 25 neurons per layer. It constructs the likelihood of a point using the three constraints on the Higgs mass, DM relic density and muon $g-2$, i.e.,
\begin{align*}
&m_{h^0}=125\pm 2~\text{GeV},\\
&\Omega h^2<0.126, \\
&\Delta a_{\mu}=(2.87\pm0.97)\times 10^{-9}.
\end{align*}
The ANN first generates a set of points using the SUGRA input parameters which are used to train the neutral network based on the constructed likelihood function. The input parameters are $m_0$, $A_0$, $m_1$, $m_2$, $m_3$ and  $\tan\beta$ where $m_0$ is the universal scalar mass, $A_0$ is the universal trilinear coupling, $m_1,m_2,m_3$ are the $\mathsf {U(1),  SU(2),  SU(3)}$ gaugino masses all at the GUT scale and $\tan\beta= \langle H_u\rangle/\langle H_d\rangle$ where
 $H_u$ gives mass to the up quarks and $H_d$ gives mass to the down quarks and the charged leptons. We notice that the ANN predicts a particle spectrum consistent with $\tilde g$SUGRA where the colored sparticles are heavy and the sleptons, staus and electroweakinos are lighter. Generating the sparticle spectrum requires evolving the renormalization group equations (RGEs) and for this we use \code{SPheno-4.0.4}~\cite{Porod:2003um,Porod:2011nf} which implements two-loop MSSM RGEs and three-loop SM RGEs while taking into account SUSY threshold effects at the one-loop level. The larger SUSY scale makes it necessary to employ a two-scale matching condition at the electroweak and SUSY scales~\cite{Staub:2017jnp} thereby improving the calculations of the Higgs boson mass and of the sparticle spectrum. The bottom quark mass and $\alpha_S$ (the fine structure constant for the $\mathsf{SU(3)_C}$)
are run up to the scale of the  $Z$ boson mass, $M_Z$,  using four-loop RGEs in the $\overline{\rm MS}$ scheme while for the top quark, the evolution starts at the pole mass and the $\overline{\rm MS}$ mass is computed by running down to the $M_Z$ scale including two-loop QCD corrections.
The tau mass is calculated at $M_Z$ including one-loop electroweak corrections. The calculation of the $\overline{\rm MS}$ Yukawas at the electroweak scale involves the first matching conditions to include SM thresholds. Those couplings are then run using 3-loop SM RGEs to $M_{\rm SUSY}$ where the second matching takes place to include SUSY thresholds at the one-loop level and a shift is made to the $\overline{\rm DR}$ scheme. The 2-loop MSSM RGEs of the $\overline{\rm DR}$ Yukawas and gauge couplings are then run to the GUT scale where the soft SUSY breaking boundary conditions are applied.
The obtained set of points are then passed to \code{Lilith}~\cite{Bernon:2015hsa,Kraml:2019sis}, \code{HiggsSignals}~\cite{Bechtle:2013xfa} and \code{HiggsBounds}~\cite{Bechtle:2020pkv} to check the Higgs sector constraints as well as \code{SModelS}~\cite{Khosa:2020zar,Kraml:2013mwa,Kraml:2014sna} to check the LHC constraints. Furthermore, \code{micrOMEGAs-5.2.7}~\cite{Barducci:2016pcb} has a module which we use to check the constraints from DM direct detection experiments.

We discuss now the results of our analysis. In Table~\ref{tab1} we give an analysis of the VEVs of the
heavy fields that enter in the GUT symmetry breaking for a range of GUT parameters $\eta, \lambda$,
$M^{126}$ and $M^{210}$ where the VEVs are in general complex. The VEVs are obtained by
solving the spontaneous symmetry breaking equations using $W_{\rm GUT}$. Using the VEVs
of Table~\ref{tab1}, one solves for the Higgs doublet mass matrix using a range of $a, b_1, b_2, c, \bar c$
that appear in $W_{\rm DT}$. The diagonalization of the Higgs mass matrix allows us to identify
the linear combination of the Higgs doublet fields which are massless and correspond to the
pair of  MSSM Higgs.

\begin{table}[H]
\begin{center}
\resizebox{\linewidth}{!}{\begin{tabulary}{\linewidth}{|l|cccccccc|}
\hline\hline\rule{0pt}{3ex}
Model&{$\eta$}&{$\lambda$}& $M^{126}$ & $M^{210}$ &$\mathcal V_{1_{_{{210}}}}$&
 $\mathcal V_{24_{_{{210}}}}$&$\mathcal V_{75_{_{{210}}}} $&$\mathcal V_{1_{_{{126}}}} $ \\
\hline\rule{0pt}{3ex}
\!\!(a) & 2.22 & 1.96 & $5.73\times 10^{17}$ & $1.14\times 10^{15}$ & $2.00\times 10^{18}$ & $(-4.00+\imath0.42)\times 10^{18}$ & $(-8.97+\imath0.38)\times 10^{18}$ & $(2.82-\imath0.09 i)\times 10^{17}$ \\
(b) & 2.85 & 2.31 & $8.10\times 10^{17}$ & $2.00\times 10^{16}$ & $2.20\times 10^{18}$ & $-4.01\times 10^{18}$ & $-2.08\times 10^{18}$ & $\imath 2.93\times 10^{18}$ \\
(c) & 3.00 & 2.88 & $4.27\times 10^{17}$ & $2.12\times 10^{15}$ & $1.10\times 10^{18}$ & $(-2.19+\imath0.36)\times 10^{18}$ & $(-4.94+\imath0.32)\times 10^{18}$ & $(2.45-\imath 0.12)\times 10^{17}$ \\
(d) & 2.62 & 0.63 & $4.31\times 10^{17}$ & $4.01\times 10^{15}$ & $1.28\times 10^{18}$ & $-2.27\times 10^{18}$ & $-1.15\times 10^{18}$ & $\imath 9.20\times 10^{17}$\\
(e) & 1.37 & 2.61 &$5.67\times 10^{17}$ & $1.41\times 10^{16}$ & $3.20\times 10^{18}$ & $(-6.32+\imath1.65)\times 10^{18}$ & $(-1.45+\imath0.15)\times 10^{19}$ & $(1.60-\imath0.12)\times 10^{18}$ \\
(f) & 1.11 & 2.51 & $3.03\times 10^{17}$ & $1.44\times 10^{16}$ & $2.12\times 10^{18}$ & $(-4.16+\imath1.39)\times 10^{18}$ & $(-9.65+\imath1.27)\times 10^{18}$ & $(1.46-\imath0.14)\times 10^{18}$\\
(g) & 2.24 & 0.90 & $4.04\times 10^{17}$ & $2.89\times 10^{15}$ & $1.40\times 10^{18}$ & $-2.65\times 10^{18}$ & $-1.42\times 10^{18}$ & $\imath 1.33\times 10^{18}$ \\
(h) & 2.98 & 2.71 &$5.09\times 10^{17}$ & $1.79\times 10^{16}$ & $1.32\times 10^{18}$ & $(-2.54+\imath1.19)\times 10^{18}$ & $(-6.09+\imath1.11)\times 10^{18}$ & $(7.83-\imath 1.04)\times 10^{17}$  \\
(i) & 2.07 & 1.13 &$3.11\times 10^{17}$ & $1.21\times 10^{16}$ & $1.16\times 10^{18}$ & $-1.86\times 10^{18}$ & $-8.71\times 10^{17}$ & $\imath 1.21\times 10^{18}$ \\
(j) & 2.99 & 0.39 &$6.61\times 10^{17}$ & $1.88\times 10^{16}$ & $1.71\times 10^{18}$ & $-1.58\times 10^{18}$ & $-4.80\times 10^{17}$ & $\imath 7.54\times 10^{17}$ \\
\hline\hline
\end{tabulary}}
\caption{{\small A numerical estimate of the VEVs of the Standard Model singlets in $\mathsf{210}$, $\mathsf{126}$ and  $\mathsf{\overline{126}}$-plets arising in the spontaneous breaking of the $\mathsf{SO(10)}$ GUT gauge symmetry under the assumption $\mathcal V_{1_{_{{126}}}}=\mathcal V_{1_{_{\overline{126}}}}$. All VEVs and masses are in GeV.}}
\label{tab1}
\end{center}
\end{table}

The diagonalization also allows for computation of non-vanishing elements
of the $U$ and $V$ matrices that connect to the light Higgs. These are the matrix  elements $U_{d_{11}}$,~
$U_{d_{21}},~ U_{d_{31}},~ U_{d_{61}}$ and the matrix elements $V_{d_{11}}$,~ $V_{d_{21}},~ V_{d_{31}},~ V_{d_{61}}$.
They are listed in Tables~\ref{tab2} and~\ref{tab3}.  In Table~\ref{tab4} we give a list of
parameters that enter in the cubic couplings $W_3$ and in the quartic couplings $W_4$.
In Table~\ref{tab5} we give the computations of the contributions of the cubic couplings, the quartic
couplings and their sum for $b,t,\tau$ for the model points of Table~\ref{tab1}.
 Computation of $b,t,\tau$ masses using the analysis of Table~\ref{tab5} as boundary conditions at
 the GUT scale and using RG evolution down to the electroweak scale is given in Table~\ref{tab6}.
An analysis of the Higgs boson mass, the light sparticle masses, the dark matter relic density and
of the supersymmetric correction to the muon anomaly is given in Table~\ref{tab7}. A comparison between Table~\ref{tab6} and Table~\ref{tab7} shows that one has a unification of Yukawas and a $g-2$
anomaly consistent with the Fermilab result of Eq.~(\ref{diff}).
One may note  that the dark matter relic density is not fully saturated by
the model points of Table~\ref{tab7}. This implies that the dark matter may likely be multicomponent
which includes other forms of dark matter, such as dark fermions of the hidden sector~\cite{Feldman:2010wy,Feldman:2011ms,Aboubrahim:2019mxn}
or possibly a dark photon~\cite{Aboubrahim:2021ycj} or an axion~\cite{Baer:2018rhs,Halverson:2017deq}.

\begin{table}[H]
\begin{center}
\resizebox{\linewidth}{!}{
\begin{tabulary}{\linewidth}{|l|ccccccccc|}
\hline\hline\rule{0pt}{3ex}
{Model}&{$a$}&{$b_{1}$}&{$b_{2}$}&{$c$}&{$\bar{c}$}& $U_{d{_{11}}}$& ${U_{d{_{21}}}}$& ${U_{d{_{31}}}}$& ${U_{d{_{61}}}}$ \\
\hline\rule{0pt}{3ex}
\!\!(a) & 0.22 & 1.86 & 1.14 & 1.46 & 0.18 & $-0.034+\imath0.051$ & $0.298+\imath0.285$ & $0.231+\imath0.351$ & $-0.495-\imath0.636$ \\
(b) & 2.03 & 2.50 & 2.70 & 0.81 & 1.15 & $-0.040-\imath0.015$ & $0.082+\imath0.030$ & $0.185+\imath0.067$ & $-0.917-\imath0.333$ \\
(c) & 1.70 & 2.95 & 1.21 & 0.21 & 2.74 & $-0.163-\imath0.009$ & $0.381+\imath0.080$ & $-0.098+\imath0.409$ & $0.091-\imath0.798$ \\
(d) & 0.27 & 2.55 & 2.16 & 0.51 & 2.65 & $0.487+\imath0.003$ & $-0.600-\imath0.003$ & $-0.121-\imath0.001$ & $0.623+\imath0.003$ \\
(e) & 2.33 & 1.65 & 1.04 & 0.08 & 2.95 & $-0.101+\imath0.165$ & $0.185-\imath0.249$ & $0.377+\imath0.213$ & $-0.782-\imath0.259$ \\
(f) & 0.16 & 1.41 & 1.53 & 0.40 & 2.46 & $-0.715-\imath0.001$ & $0.661+\imath0.023$ & $0.015+\imath0.106$ & $-0.075-\imath0.187$ \\
(g) & 0.51 & 2.90 & 1.08 & 0.19 & 1.37 & $0.096+\imath0.138$ & $-0.272-\imath0.390$ & $-0.103-\imath0.148$ & $0.482+\imath0.693$ \\
(h) & 2.52 & 2.91 & 0.21 & 0.25 & 2.99 & $-0.067+\imath0.047$ & $0.850-\imath0.440$ & $0.101+\imath0.085$ & $-0.230-\imath0.087$ \\
(i) & 1.57 & 1.38 & 2.45 & 0.75 & 1.41 & $0.067-\imath0.043$ & $-0.072+\imath0.047$ & $-0.137+\imath0.089$ & $0.823-\imath0.531$ \\
(j) & 0.68 & 2.70 & 1.01 & 0.21 & 0.49 & $0.130-\imath0.003$ & $-0.361+\imath0.008$  & $-0.072+\imath0.002$ & $0.920-\imath0.021$ \\
\hline\hline
\end{tabulary}}
\caption{{\small A numerical estimate of the elements of the down Higgs zero mode eigenvector using the analysis of Table \ref{tab1}
and the couplings of Eq.~(\ref{dtsuper}).}}
\label{tab2}
\end{center}
\end{table}

\begin{table}[H]
\begin{center}
\resizebox{\linewidth}{!}{
\begin{tabular}{|l|ccccccccc|}
\hline\hline\rule{0pt}{3ex}
{Model}&{$a$}&{$b_{1}$}&{$b_{2}$}&{$c$}&{$\bar{c}$}&$V_{d{_{11}}}$&${V_{d{_{21}}}}$&${V_{d{_{31}}}}$&${V_{d{_{61}}}}$ \\
\hline\rule{0pt}{3ex}
\!\!(a) & 0.22 & 1.86 & 1.14 & 1.46 & 0.18 & $-0.273$ & $0.411+\imath0.083$ &$-0.401$  & $0.765+\imath0.062$ \\
(b) & 2.03 & 2.50 & 2.70 & 0.81 & 1.15 & $-0.091$ & $0.107$ & $-0.196$ & $0.971$ \\
(c) & 1.70 & 2.95 & 1.21 & 0.21 & 2.74 & $0.323$ & $-0.783-\imath0.010$ & $0.246$ & $-0.467-\imath0.057$ \\
(d) & 0.27 & 2.55 & 2.16 & 0.51 & 2.65 & $-0.592$ & $0.708$ & $-0.074$ & $0.378$ \\
(e) & 2.33 & 1.65 & 1.04 & 0.08 & 2.95 & $-0.357$ & $0.568+\imath0.010$ & $-0.345$ & $0.644+\imath0.127$ \\
(f) & 0.16 & 1.41 & 1.53 & 0.40 & 2.46 & $0.730$ & $-0.673-\imath0.006$ & $0.057$ & $-0.104-\imath0.026$ \\
(g) & 0.51 & 2.90 & 1.08 & 0.19 & 1.37 & $-0.276$ & $0.747$ & $-0.127$ & $0.591$ \\
(h) & 2.52 & 2.91 & 0.21 & 0.25 & 2.99 & $0.086$ & $-0.977-\imath0.055$ & $0.089$ & $-0.157-\imath0.055$ \\
(i) & 1.57 & 1.38 & 2.45 & 0.75 & 1.41& $-0.120$  & 0.095 & $-0.163$ & 0.975  \\
(j) & 0.68 & 2.70 & 1.01 & 0.21 & 0.49 & $-0.046$  & 0.163  & $-0.077$  & 0.983 \\
\hline\hline
\end{tabular}}
\caption{{\small A numerical estimate of the elements of the up Higgs zero mode eigenvector using the analysis of Table \ref{tab1} and the couplings of Eq.~(\ref{dtsuper}).}}
\label{tab3}
\end{center}
\end{table}

\begin{table}[H]
\begin{center}
\begin{tabulary}{1.20\textwidth}{|l|CCCC|}
\hline\hline\rule{0pt}{3ex}
Model & $f^{(1)}$ & $f^{(2)}$ & $f^{(3)}$ & $f^{10_r}$\\
\hline\rule{0pt}{3ex}
\!\!(a) & 0.16 & 0.24 & 0.03 & (0.17, 0.23)\\
(b) & 0.12 & 0.10 & 0.12  & (0.30, 1.04) \\
(c) & 0.40 & 0.08 & 0.08  & (2.36, 1.06)\\
(d) & 0.68 & 0.35 & 0.22  & (0.43, 0.44)\\
(e) & 1.24 & 0.10 & 0.04  & (0.12, 0.25) \\
(f) & 1.58 & 0.63 & 0.10 & (2.03, 2.26) \\
(g) & 0.79 & 0.15 & 0.14 & (0.38, 0.24) \\
(h) & 1.55 & 0.38 & 0.22 & (1.55, 0.21)\\
(i) & 0.15 & 0.11 & 0.08 & (1.29, 2.30)\\
(j) & 0.44 & 0.09 & 0.22 & (0.52, 0.49) \\
\hline
\end{tabulary}\end{center}
\caption{{\small The GUT scale parameters in the cubic and quartic superpotentials $W_3$,  $W_4^{(1)}$, $W_4^{(2)}$ and  $W_4^{(3)}$
for the model points (a)$-$(j). The masses are in GeV.}}
\label{tab4}
\end{table}

\begin{table}[H]
\begin{center}
\begin{tabulary}{1.10\textwidth}{|l|CCCCCCCCC|}
\hline\hline\rule{0pt}{3ex}
Model & $h_t^0$ & $h_b^0$ & $h_\tau^0$ & $\delta h_t^{\rm GUT}$ &$\delta h_b^{\rm GUT}$ & $\delta h_{\tau}^{\rm GUT}$ & $h_t^{\rm GUT}$ & $h_b^{\rm GUT}$ & $h_{\tau}^{\rm GUT}$  \\
\hline\rule{0pt}{3ex}
\!\!(a) & 0.274 & 0.148 & 0.148 & 0.204 & 0.201 & 0.063 & 0.478 & 0.073 & 0.088 \\
(b) & 0.223 & 0.238 & 0.238 & 0.259 & 0.282 & 0.183 & 0.482 & 0.044 &  0.055 \\
(c) & 0.190 & 0.193 & 0.193 & 0.319 & 0.190 & 0.163 & 0.501 & 0.029 & 0.036 \\
(d) & 0.161 & 0.169 & 0.169 & 0.331 & 0.236 & 0.089 & 0.492 & 0.066 &  0.081 \\
(e) & 0.153 & 0.278 & 0.278 & 0.400 & 0.341 & 0.231 & 0.486 & 0.062 &  0.074 \\
(f) & 0.189 & 0.118 & 0.118 & 0.298 & 0.200 & 0.108 & 0.484 & 0.091 &  0.104 \\
(g) & 0.149 & 0.222 & 0.222 & 0.348 & 0.272 & 0.159 & 0.497 & 0.051 & 0.062 \\
(h) & 0.210 & 0.198 & 0.198 & 0.289 & 0.220 & 0.156 & 0.487 & 0.042 & 0.053 \\
(i) & 0.268 & 0.179 & 0.179 & 0.216 & 0.248 & 0.094 & 0.483 & 0.068 &  0.085 \\
(j) & 0.313 & 0.160 & 0.160 & 0.177 & 0.211 & 0.099 & 0.489 & 0.051 & 0.060 \\
\hline
\end{tabulary}\end{center}
\caption{
The  magnitude of the contributions to the top, bottom, and tau Yukawa couplings from cubic interactions (columns 2-4),
 from quartic interactions (columns 5-7)
 and the magnitude of their complex  sum (columns 8-10)
 at the GUT scale for the parameter set  of Table \ref{tab4}.
The Yukawa couplings are in general complex and we add the contributions of the cubic and
quartic interactions as complex numbers and exhibit only their magnitudes in the table.}
\label{tab5}
\end{table}

\begin{table}[H]
\begin{center}
\begin{tabulary}{1.10\textwidth}{|l|CCCCCC|CCC|}
\hline\hline\rule{0pt}{3ex}
Model & $m_0$ & $A_0$ & $m_1$ & $m_2$ & $m_3$ & $\tan\beta$  &$m_t$ (pole) & $\overline{m}_b(\overline{m}_b)$ & $m_{\tau}$ (pole)  \\
\hline\rule{0pt}{3ex}
\!\!(a) & 657 & -2228 & 661 & 526 & 7774 & 14.0  & 172.2 & 4.15 & 1.77682  \\
(b) & 673 & 1127 & 939 & 570 & 8833 & 8.2  & 172.2 & 4.22 & 1.77682  \\
(c) & 387 & 880 & 949 & 980 & 8118 & 5.3 & 172.8 & 4.19 & 1.77682  \\
(d) & 164 & 197 & 632 & 1539 & 6171 & 12.2  & 172.9 & 4.20 & 1.77682  \\
(e) & 416 & 339 & 740 & 416 & 4559 & 11.6  & 172.8 & 4.22 & 1.77682 \\
(f) & 688 & 1450 & 852 & 634 & 8438 & 16.8  & 172.9 & 4.22 & 1.77682  \\
(g) & 106 & 22.6 & 523 & 1309 & 5240 & 9.3  & 172.8 & 4.19 & 1.77682  \\
(h) & 206 & 603 & 842 & 1298 & 7510 & 8.0  & 172.1 & 4.15 & 1.77682 \\
(i) & 452 & 648 & 624 & 346 & 4843 & 13.1 & 172.8 & 4.20 & 1.77682 \\
(j) & 196 & -803 & 828 & 1599 & 8929 & 9.4  & 172.6 & 4.22 & 1.77682 \\
\hline
\end{tabulary}\end{center}
\caption{
The SUGRA parameters sets used for RG analysis where the boundary conditions for the Yukawas for the top, bottom, and the tau
are taken from Table \ref{tab5}. In the analysis the GUT scale ranges from $8.6\times 10^{15}$ GeV to $2.0\times 10^{16}$ GeV.}
\label{tab6}
\end{table}

\begin{table}[H]
\begin{center}
\begin{tabulary}{1.10\textwidth}{|l|CCCCCCCC|}
\hline\hline\rule{0pt}{3ex}
Model & $h^0$  & $\tilde \mu$ & $\tilde \nu_{\mu}$ & $\tilde\tau$ & $\tilde\chi^0_1$ & $\tilde\chi^{\pm}_1$ & $\Omega h^2$ & $\Delta a_{\mu}(\times 10^{-9})$ \\
\hline\rule{0pt}{3ex}
\!\!(a) & 123.3 & 459.0 & 452.6 & 270.8  & 243.1 & 323.0 & 0.103 & 2.30 \\
(b) & 125.3 & 422.8 & 415.7  & 370.4  & 337.3 &  337.6 & 0.003 & 2.14 \\
(c) & 123.3 & 427.2  & 420.5  & 379.6  & 369.8  & 707.7  & 0.125 & 1.91 \\
(d) & 123.9 & 856.4 & 852.4 & 243.5 & 240.1 & 1227  & 0.016 & 1.94\\
(e) & 123.8 & 361.0 & 352.6 & 282.0 & 272.7 & 272.9 & 0.002 & 1.98 \\
(f) & 123.0 & 508.1 & 502.3 & 331.9 & 324.2 & 404.3 & 0.004 & 2.11 \\
(g) & 123.4 & 722.8 & 718.2  & 206.5 & 195.5 & 1038.4 &  0.103 & 2.57 \\
(h) & 124.5 & 628.7 & 623.6  & 338.3 & 326.8 & 998.4  & 0.082 & 1.94 \\
(i) & 123.7 & 346.8 & 338.0 & 240.3 & 205.6 & 205.8 & 0.001 & 2.67 \\
(j) & 123.5 & 774.1 & 769.8 & 319.1 & 314.7 & 1247 & 0.016 & 2.59 \\
\hline
\end{tabulary}\end{center}
\caption{Low scale SUSY mass spectrum showing the Higgs boson, the smuon, the muon sneutrino, the stau and the light electroweakino masses and the LSP relic density for the benchmarks  of Table~\ref{tab6}. Also shown is $\Delta a_{\mu}$.}
\label{tab7}
\end{table}

\begin{figure}[H]
 \centering
 \includegraphics[width=0.49\textwidth]{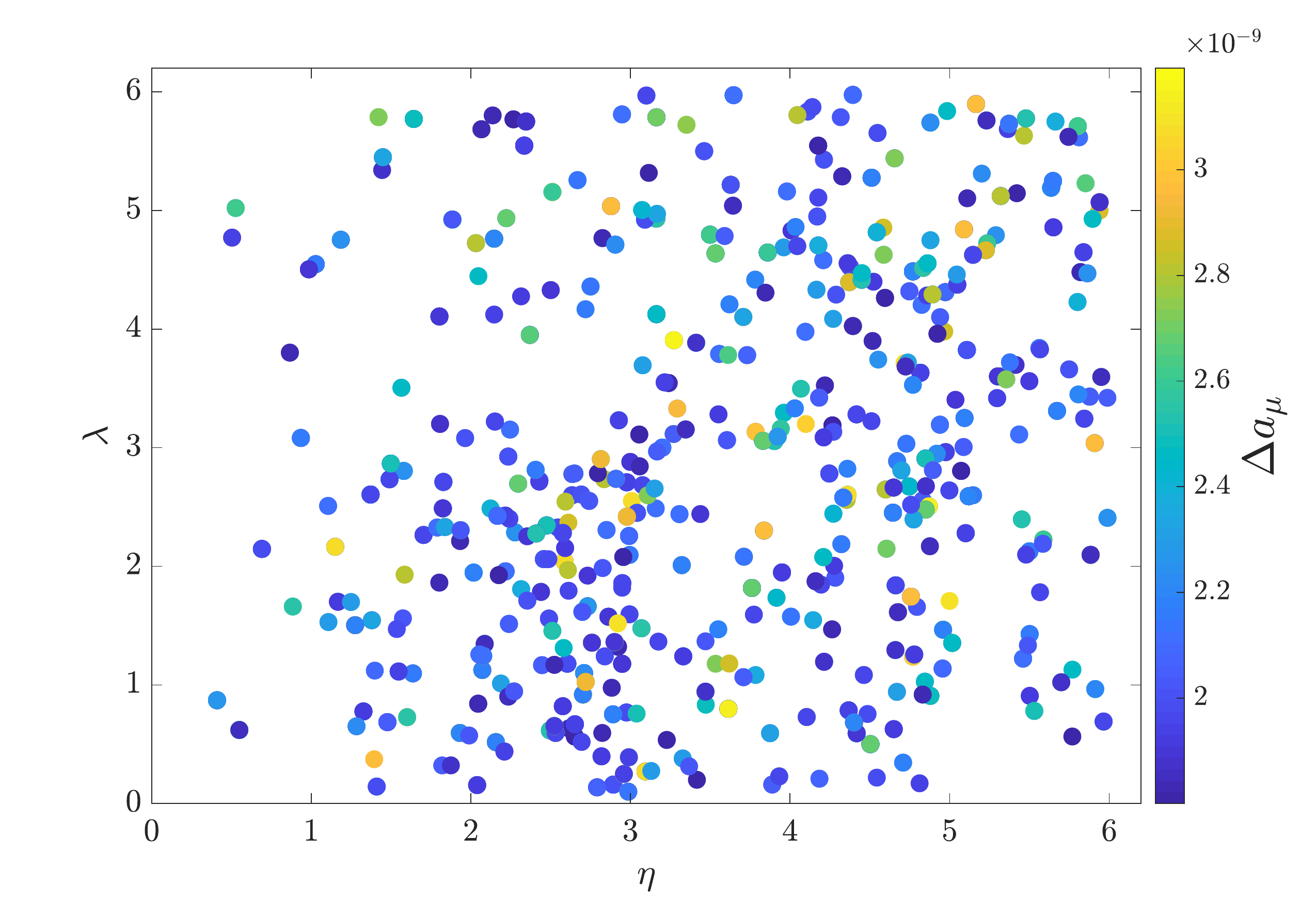}
 \includegraphics[width=0.49\textwidth]{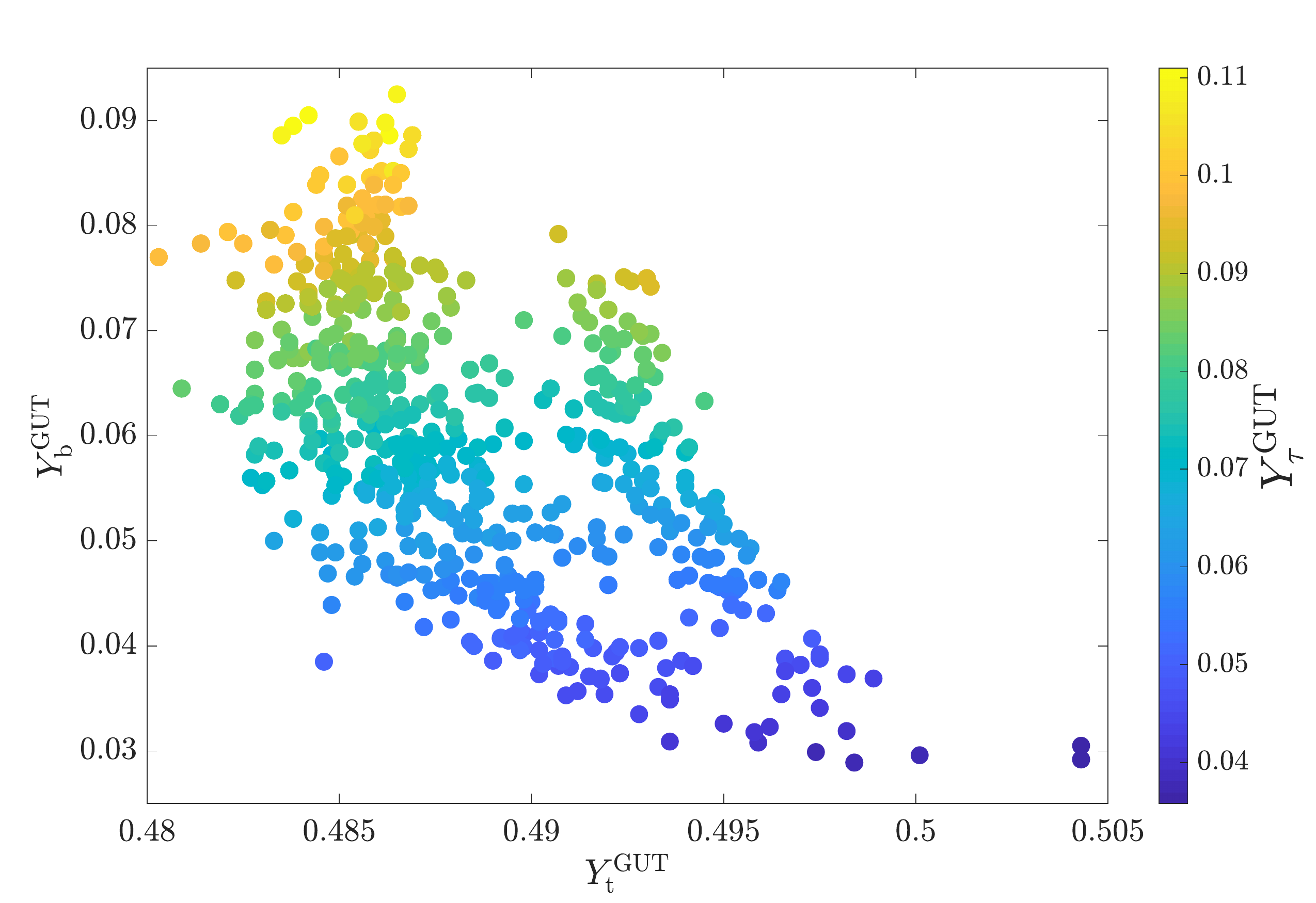}
   \caption{Scatter plots resulting from the scan of input parameters from $\mathsf{SO(10)}$. The left panel shows the parameters $\eta$ and $\lambda$ with the color axis being the muon $g-2$. The right panel shows the top, bottom and tau Yukawa couplings at the GUT scale. }
	\label{fig1}
\end{figure}

A scan on the parameter space using the GUT scale input of $\mathsf{SO(10)}$ results in a larger set of points than those presented in Tables~\ref{tab1}$-$\ref{tab5}. The range of values the input parameters take are: $0.5<\eta,\lambda<6.0$, $0.1<a,b_1,b_2,c,\bar{c}<3.0$, $1\times10^{16}<M^{126}<9.5\times 10^{17}$, $1\times10^{15}<M^{210}<3.5\times 10^{16}$, $0.01<f^{(1)},f^{(2)},f^{(3)}<4.0$ and $0.1<f^{10_r}<5.5$. The result of the scan is shown in Fig.~\ref{fig1}. The left panel is a scatter plot in the variables $\eta$ and $\lambda$ with the muon $g-2$ shown on the color axis consistent with \amu. The right panel shows a scatter plot in the top, bottom and tau Yukawa couplings at the GUT scale. The set of points in the scatter plot is consistent with experimental constraints and the evolution of the GUT scale Yukawas to the electroweak scale produces the correct top, bottom and tau masses within experimental uncertainties.

\section{Sparticle hierarchies and signal region analysis \label{sec4}}

The set of data points retained after satisfying the constraints from the Higgs sector, the DM relic density, dark matter direct detection and the LHC is further processed and points consistent with Yukawa coupling unification are kept. We observe that the spectrum consisting of light electroweakinos, sleptons (selectron and smuons) and staus belong to three cases of mass hierarchy.
\paragraph{Case 1:} The electroweakinos, $\tilde\chi^0_2,\tilde\chi^{\pm}_1$ are almost degenerate, with the stau being the next-to-lightest supersymmetric particle (NLSP). The mass hierarchy here is $$m_{\tilde\tau_1}<m_{\widetilde{\rm EW}}<m_{\tilde\ell},$$ where $\widetilde{\rm EW}=(\tilde\chi^0_2,\tilde\chi^{\pm}_1)$ and $\tilde\ell$ represents the sleptons.
\paragraph{Case 2:} In this category, one of the electroweakinos ($\tilde\chi^0_2$ or $\tilde\chi^{\pm}_1$) is the NLSP and the hierarchy reads $$m_{\widetilde{\rm EW}}<m_{\tilde\tau_1}<m_{\tilde\ell}\,.$$ Here we distinguish two subcategories (I) and (II) where
\begin{align*}
&m_{\tilde\chi^{\pm}_1}<m_{\tilde\chi^0_2}<m_{\tilde\tau_1}~~~\text{(I)}, \\
&m_{\tilde\chi^{\pm}_1}<m_{\tilde\tau_1}<m_{\tilde\chi^0_2}~~~\text{(II)}.
\end{align*}
\paragraph{Case 3:} The last category also includes stau as the NLSP but the electroweakino and slepton hierarchy is inverted, i.e.,
$$m_{\tilde\tau_1}<m_{\tilde\ell}<m_{\widetilde{\rm EW}}.$$

\begin{figure}[H]
 \centering
 \includegraphics[width=0.7\textwidth]{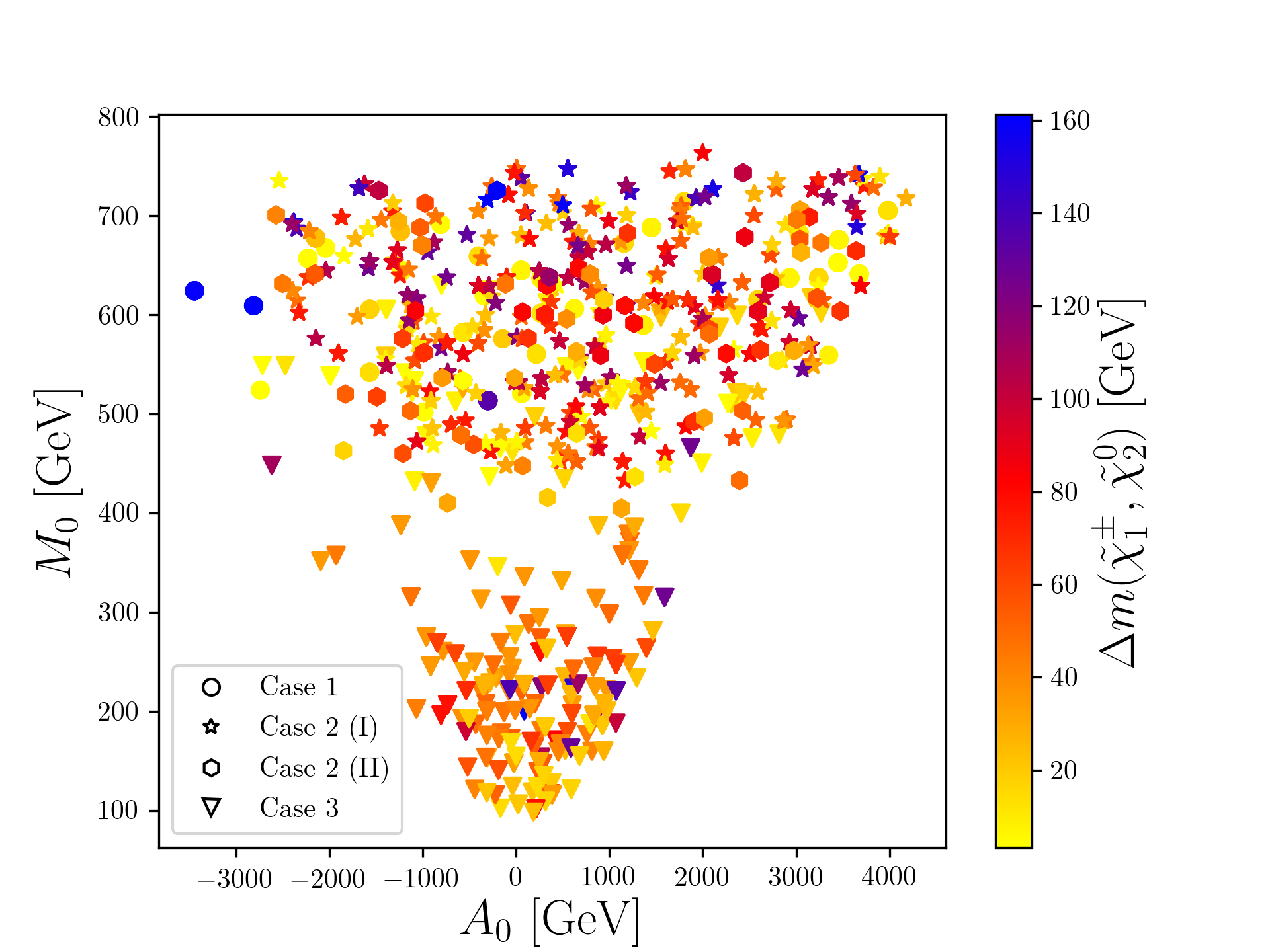}
 \caption{A scatter plot in the $M_0$-$A_0$ plane showing the three cases (and subcases) with the chargino-second neutralino mass gap shown on the color axis. }
	\label{fig2}
\end{figure}

Benchmarks (a), (f) belong to Case 1, while (b), (e) and (i) belong to Case 2 and (c), (d), (g), (h) and (j) belong to Case 3. Fig.~\ref{fig2} shows the obtained data set categorized according to the above three cases.

\begin{figure}[H]
 \centering
 \includegraphics[width=0.49\textwidth]{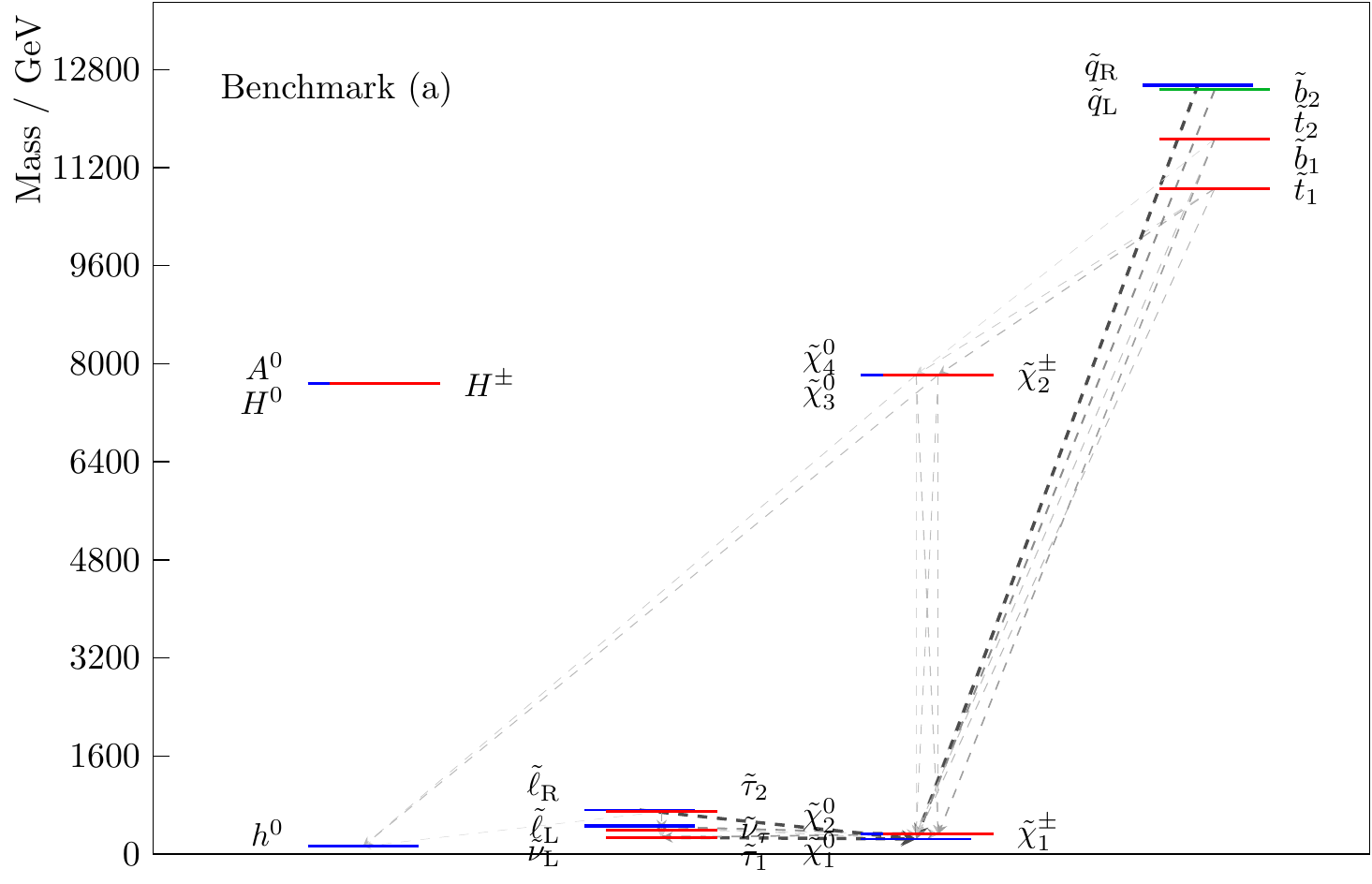}
 \includegraphics[width=0.49\textwidth]{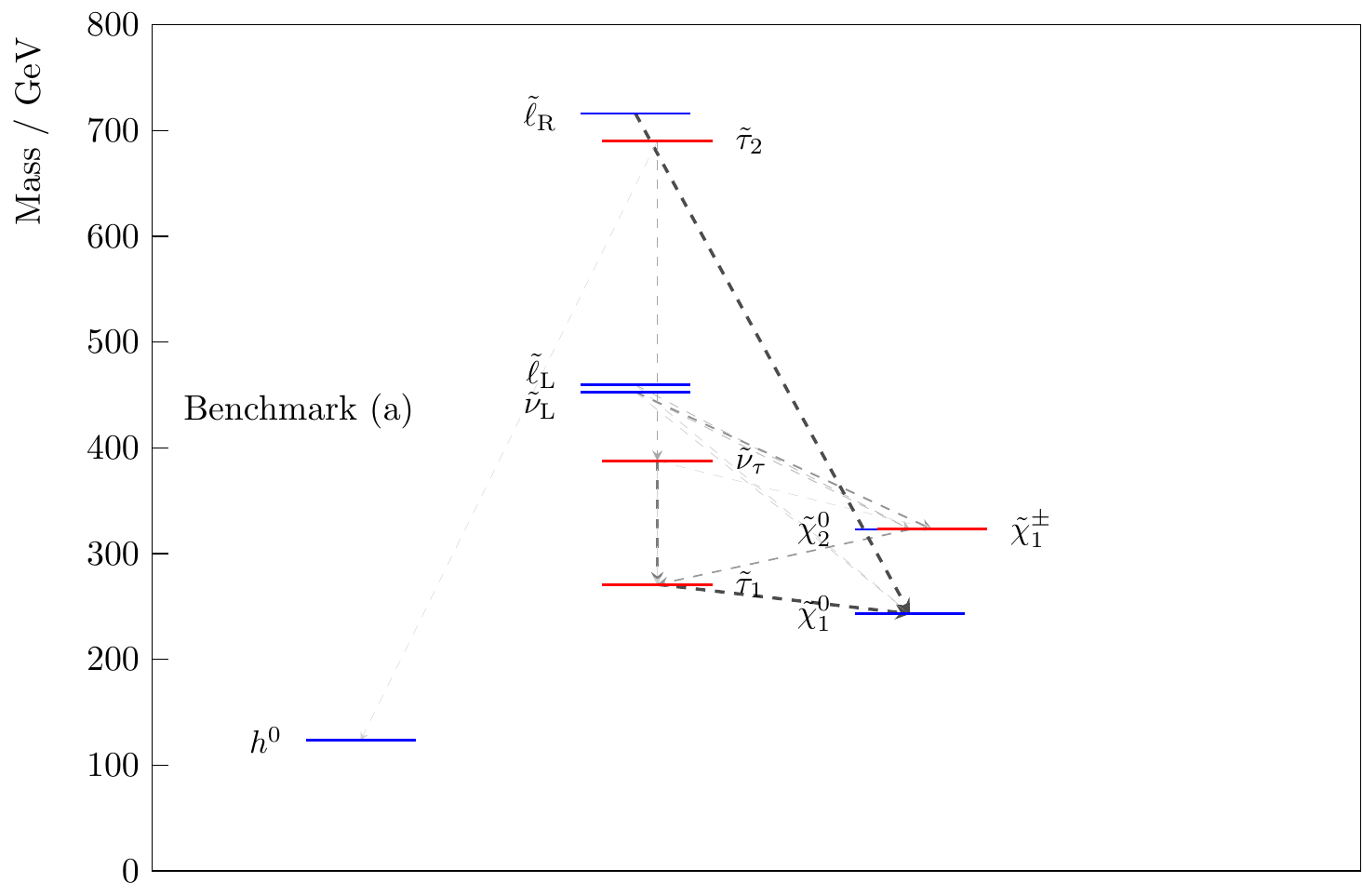}\\
 \includegraphics[width=0.49\textwidth]{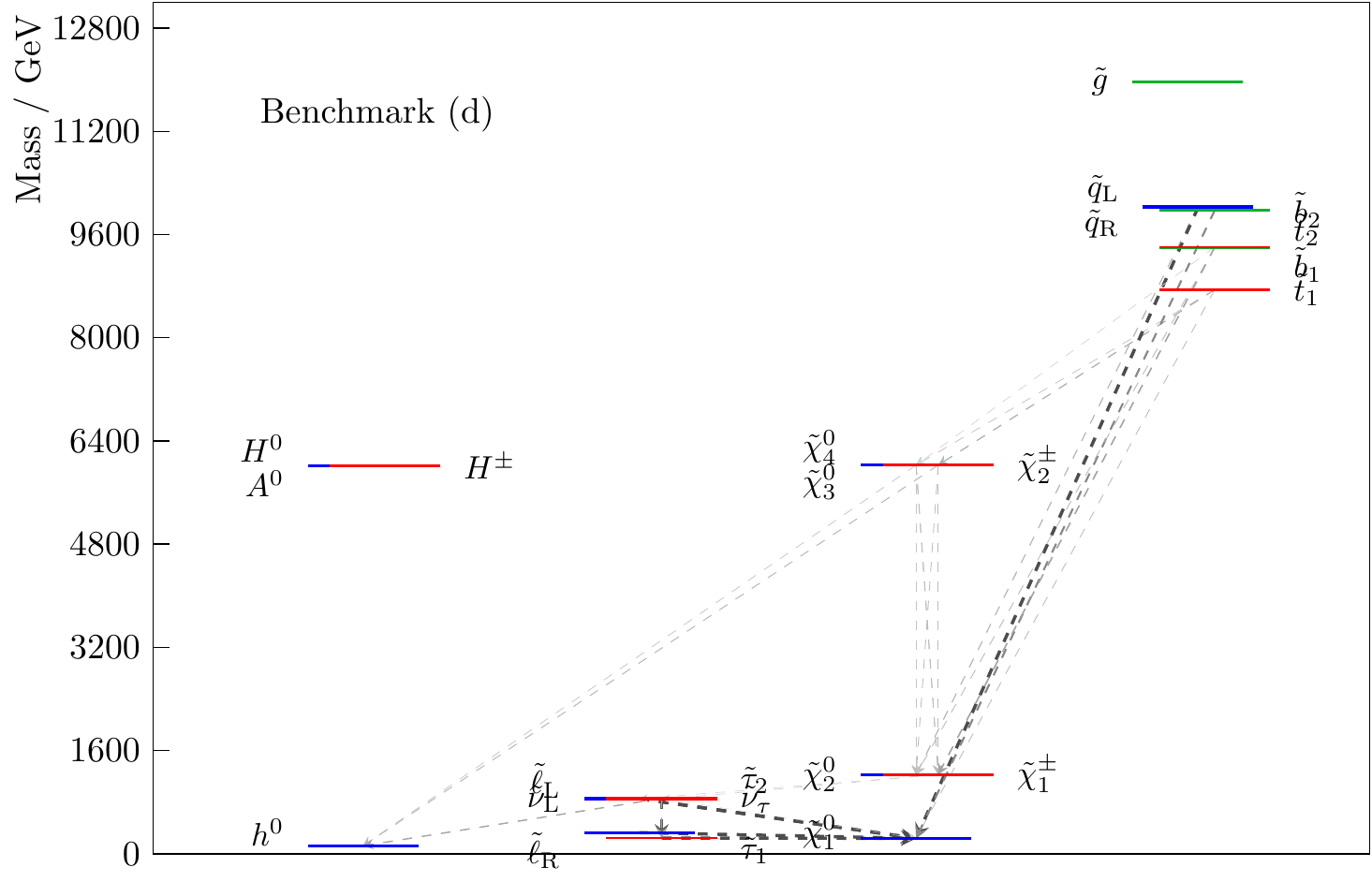}
 \includegraphics[width=0.49\textwidth]{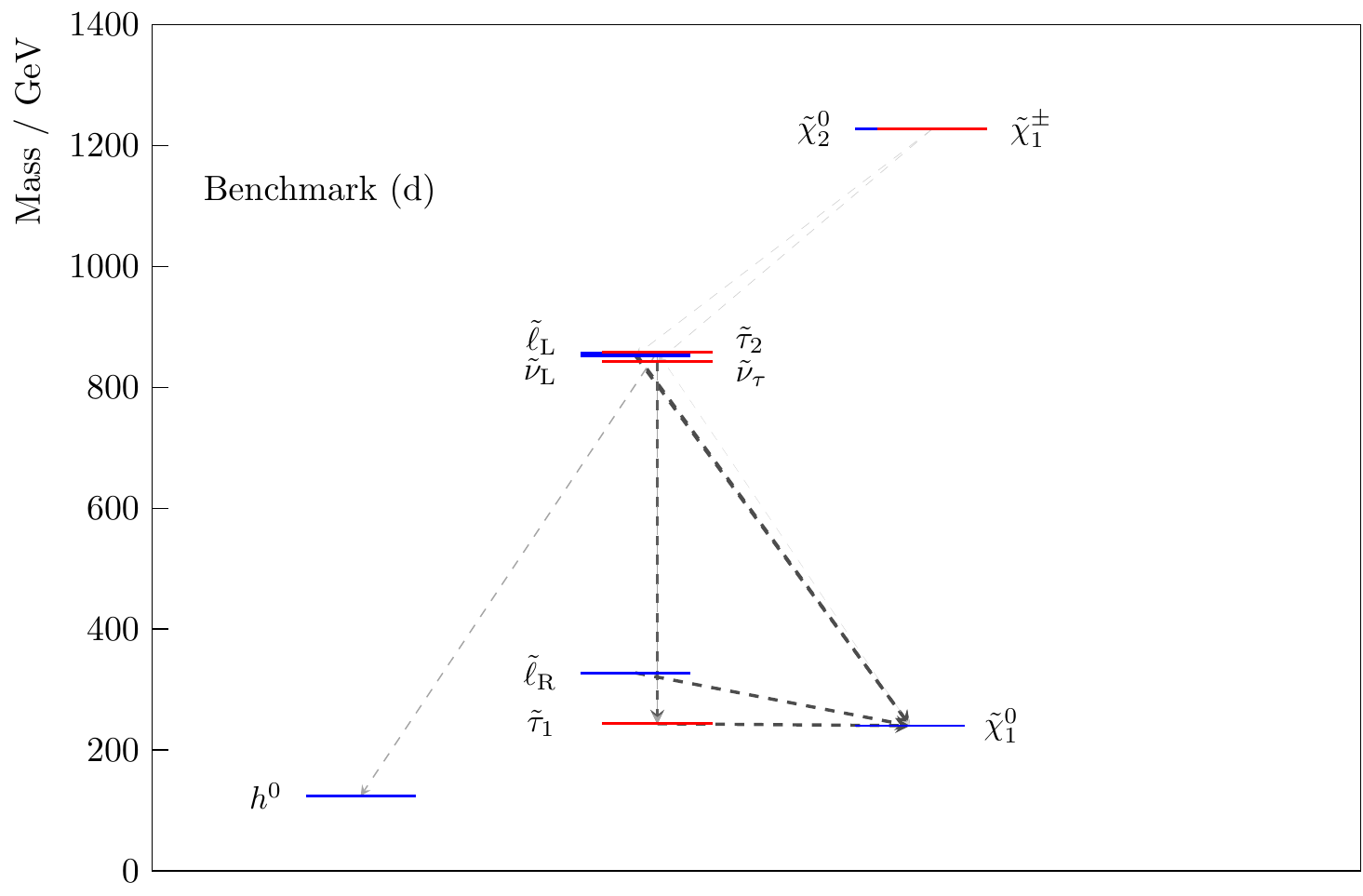}
   \caption{A display of the particle spectrum using \code{PySLHA}~\cite{Buckley:2013jua} for benchmarks (a) (upper panels) and (d) (lower panels). The left panels represent the spectrum up to 13 TeV while the right panels give the low-lying masses of the spectrum. }
	\label{fig3}
\end{figure}

An illustration of such a complex spectrum is given in Fig.~\ref{fig3}. The upper panels correspond to benchmark (a) while the lower ones are for (d). Cascade decays are common in high scale models which, unlike simplified models considered by ATLAS and CMS, produce more complicated event topology. Thus, for slepton pair production, analyses by ATLAS~\cite{Aad:2019vnb,Aad:2019qnd} and CMS~\cite{Sirunyan:2018nwe,Sirunyan:2020eab} consider a 100\% branching ratio of $\tilde\ell\to\ell\tilde\chi^0_1$ which can happen in spectra belonging to Case 3. However, Cases 1 and 2 do not necessarily abide by this and one can get several decay channels making the final states more complicated.

In the next section, we select a set of benchmarks belonging to the three cases discussed above. We study slepton pair production and decay at HL-LHC and HE-LHC. We design a set of signal regions to target the rich final states corresponding to the three cases of mass hierarchies. For earlier works on SUSY discovery at HL-LHC and HE-LHC, see Refs.~\cite{Aboubrahim:2018bil,Aboubrahim:2018tpf} and the CERN yellow reports~\cite{Cepeda:2019klc,CidVidal:2018eel}.

\subsection{Slepton pair production and event simulation at the LHC}

The pair production cross section of sleptons (selectrons and smuons) is proportional to the electron and muon Yukawa coupling which means that those cross sections are small compared to staus and electroweak gauginos. For our LHC analysis, we select six of the ten benchmarks shown in Table~\ref{tab6} corresponding to sleptons in the mass range of $\sim  350$ GeV to $\sim 850$ GeV. The production cross sections of the slepton pairs at 14 TeV and 27 TeV are calculated at the aNNLO+NNLL accuracy using~\code{Resummino-3.0}~\cite{Debove:2011xj,Fuks:2013vua} and the five-flavor NNPDF23NLO PDF set. The results, arranged in decreasing order of cross section, are shown in Table~\ref{tab8}. Also shown are the different branching ratios of sleptons but for brevity we do not exhibit the branching ratios of $\tilde\chi^0_2$ and $\tilde\chi^{\pm}_1$ for benchmarks (b), (f) and (i). To have an idea of the decay channels involved, one can examine the right panel of Fig.~\ref{fig3} which shows the low-lying spectrum of benchmark (a). Since (a) and (f) both belong to Case 1, one can have an idea of the different decay channels of $\tilde\chi^0_2$ and $\tilde\chi^{\pm}_1$ which involve the stau. This leads to a tau-enriched final state.

\begin{table}[H]
\centering
\begin{tabular}{|ccc|cc|ccc|}
\hline\hline
Model & \multicolumn{2}{c|}{$\sigma(pp\rightarrow \tilde e_L\,\tilde e_L)$}& \multicolumn{2}{c|}{$\sigma(pp\rightarrow \tilde\mu_L\,\tilde\mu_L)$} & \multicolumn{3}{c|}{Branching ratios}  \\
\hline
&14 TeV & 27 TeV & 14 TeV & 27 TeV & $\tilde\ell_{L}\to \ell\tilde\chi^0_1$ & $\tilde\ell_{L}\to \ell\tilde\chi^0_2$ & $\tilde\ell_{L}\to \nu_{\ell}\tilde\chi^{\pm}_1$ \\
\hline \rule{0pt}{3ex}
\!\!(i) & 2.896 & 9.633  & 2.909 & 9.673 & 31\% & 6\% & 63\% \\
(b) & 1.242 & 4.590 & 1.244 & 4.598 & 31\% & 6\% & 63\%   \\
 (f) & 0.541 &  2.252 & 0.543  & 2.262 & 22\% & 26\% & 52\%   \\
  (h) & 0.194 &  0.958  & 0.194  & 0.957 & 100\% & - & - \\
   (g) & 0.094 & 0.533 & 0.094 & 0.533 & 100\% & - & - \\
 (d) & 0.037 & 0.253 & 0.037 & 0.253 & 100\% & - & - \\
  \hline
\end{tabular}
\caption{The aNNLO+NNLL pair production cross-sections, in fb, of sleptons at $\sqrt{s}=14$ TeV and at $\sqrt{s}=27$ TeV for benchmarks (b), (d) and (f)$-$(i) of Table~\ref{tab1} arranged in decreasing order of production cross sections. Also shown are the slepton branching ratios to electroweakinos and leptons.}
\label{tab8}
\end{table}

The final states which make up our signal region (SR) involve two same flavor and opposite sign (SFOS) leptons with missing transverse energy (MET). We also require at least two jets (N $\geq 2$) which can be used to form kinematic variables that are effective for jetty final states. We call the signal region SR-2$\ell$Nj. For such final states, the dominant SM backgrounds are from diboson production, $Z/\gamma+$jets, dilepton production from off-shell vector bosons ($V^*\rightarrow\ell\ell$), $t\bar{t}$ and $t+W/Z$. The subdominant backgrounds are Higgs production via gluon fusion ($ggF$ H) and vector boson fusion (VBF).   The simulation of the signal and background events is performed at LO with \code{MadGraph5\_aMC@NLO-3.1.0} interfaced to \code{LHAPDF}~\cite{Buckley:2014ana} using the NNPDF30LO PDF set. Up to two hard jets are added at generator level. The parton level events are passed to \code{PYTHIA8}~\cite{Sjostrand:2014zea} for showering and hadronization using a five-flavor matching scheme in order to avoid double counting of jets. For the signal events, the matching/merging scale is set at one-fourth the mass of the pair produced sleptons. Additional jets from ISR and FSR are added to the signal and background events. Jets are clustered with \code{FastJet}~\cite{Cacciari:2011ma} using the anti-$k_t$ algorithm~\cite{Cacciari:2008gp} with jet radius $R=0.4$. \code{DELPHES-3.4.2}~\cite{deFavereau:2013fsa} is then employed for detector simulation and event reconstruction using the HL-LHC and HE-LHC card. The SM backgrounds are scaled to their relevant NLO cross sections while aNNLO+NNLL cross sections are used for the signal events.

\subsection{Event selection}

The selected SFOS leptons must have a leading and subleading  transverse momenta $p_T>15$ GeV for electrons and $p_T>10$ GeV for muons with $|\eta|<2.5$. Each event should contain at least two non-b-tagged jets with the leading $p_T>20$ GeV in the $|\eta|<2.4$ region and a missing transverse energy $E^{\rm miss}_T>70$ GeV. Despite the specific preselection criteria, the analysis cuts used for the six benchmarks cannot be the same. This is due to the rich final states involved. To help us discriminate the signal from the background events, we use a set of kinematic variables along with a deep neural network (DNN) which is trained and tested on two independent sets of signal and background samples. We list the kinematic variables that enter in the training of the DNN:

\begin{enumerate}

\item $E^{\rm miss}_T$: the missing transverse energy in the event. It is usually high for the signal due to the presence of neutralinos.

\item The transverse momentum of the leading non-b tagged jets, $p_T(j_1)$. Rejecting b-tagged jets reduces the $t\bar{t}$ background.

\item The transverse momentum of the leading lepton (electron or muon), $p_T(\ell_1)$.

\item $M_{\rm T2}$, the stransverse mass~\cite{Lester:1999tx, Barr:2003rg, Lester:2014yga} of the leading and subleading leptons
\begin{equation}
    M_{\rm T2}=\min\left[\max\left(m_{\rm T}(\mathbf{p}_{\rm T}^{\ell_1},\mathbf{q}_{\rm T}),
    m_{\rm T}(\mathbf{p}_{\rm T}^{\ell_2},\,\mathbf{p}_{\rm T}^{\text{miss}}-
    \mathbf{q}_{\rm T})\right)\right],
    \label{mt2}
\end{equation}
where $\mathbf{q}_{\rm T}$ is an arbitrary vector chosen to find the appropriate minimum and the transverse mass $m_T$ is given by
\begin{equation}
    m_{\rm T}(\mathbf{p}_{\rm T1},\mathbf{p}_{\rm T2})=
    \sqrt{2(p_{\rm T1}\,p_{\rm T2}-\mathbf{p}_{\rm T1}\cdot\mathbf{p}_{\rm T2})}.
\end{equation}

\item The quantity $M^{\rm min}_{\rm T}$ defined as $M^{\rm min}_{\rm T}=\text{min}[m_{\rm T}(\textbf{p}_{\rm T}^{\ell_1},\textbf{p}^{\rm miss}_{\rm T}),m_{\rm T}(\textbf{p}_{\rm T}^{\ell_2},\textbf{p}^{\rm miss}_{\rm T})]$.
The variables $M_{\rm T2}$ and $M^{\rm min}_{\rm T}$ are effective when dealing with large MET in the final state.

\item The dilepton invariant mass, $m_{\ell\ell}$, helps in rejecting the diboson background with a peak near the $Z$ boson mass which can be done by setting $m_{\ell\ell}>100$ GeV.

\item The opening angle between the MET system and the dilepton system, $\Delta\phi(\textbf{p}_{\rm T}^{\ell},\textbf{p}^{\rm miss}_{\rm T})$, where $\textbf{p}_{\rm T}^{\ell}=\textbf{p}_{\rm T}^{\ell_1}+\textbf{p}_{\rm T}^{\ell_2}$.

\item The smallest opening angle between the first three leading jets in an event and the MET system, $\Delta\phi_{\rm min}(\textbf{p}_{\rm T}(j_i),\textbf{p}^{\rm miss}_{\rm T})$, where $i=1,2,3$.

\end{enumerate}

We use the DNN implementation in the `Toolkit for Multivariate Analysis' (TMVA)~\cite{Speckmayer:2010zz} framework within \code{ROOT6}~\cite{Antcheva:2011zz}. The DNN employed has three dense hidden layers with 128 neurons per layer and $\tanh$ as an activation function to define the output neurons given the input values. The DNN trains on the signal and background events using the above set of kinematic variables in three phases with a decreasing learning rate. After the `learning' process is over, the DNN tests the predictions on another set of signal and background samples. Despite having one background set, the training and testing must be done every time a signal sample is used, i.e., six times in our case. During the testing stage, the DNN creates a new discriminator which is called the DNN response or the DNN score. Cuts on this new variable maximizes the signal ($S$) to background ($B$) ratio, $S/\sqrt{S+B}$.

We give in Table~\ref{tab9} the set of analysis cuts on a select number of kinematic variables along with the new `DNN response' variable. Variations in cuts are used for our six benchmarks depending on the hierarchy of the spectrum which allows us to put them in three categories with (b),(i) as the first, (f) as the second and (d),(g),(h) as the third. The values shown in parentheses are the modified cuts at 27 TeV which are essential to improving the $S/\sqrt{S+B}$ ratio.

\begin{table}[H]
\centering
\begin{tabular}{|c|ccc|}
\hline\hline
Variable & (b), (i) & (f) & (d), (g), (h)  \\
\hline \rule{0pt}{3ex}
\!\! $m_{\ell\ell}~\text{[GeV]} >$ & 136 (110) & 150  & 150 (110)   \\
 $E^{\rm miss}_T/\textbf{p}^{\ell}_{\rm T}>$ & 1.9 (2.8)  &  -  & -   \\
$\Delta\phi_{\rm min}(\textbf{p}_{\rm T}(j_i),\textbf{p}^{\rm miss}_{\rm T})~\text{[rad]} >$ & -  & 0.85 (1.5)  &  -  \\
$p_T^{\ell_2}~\text{[GeV]} >$ & -  &  -  & 190 (370)   \\
$M_{T2}~\text{[GeV]} >$ & - (140)  &  - (120)  & 200 (300)   \\
DNN response $>$   & 0.9  & 0.9 & 0.9  \\
  \hline
  $\mathcal{L}$ at 14 TeV [fb$^{-1}$] & NV, 1887 & 1262 & NV, 2074, 1738 \\
$\mathcal{L}$ at 27 TeV [fb$^{-1}$] & 2804, 1320 & 694 & 1031, 689, 1194 \\
\hline
\end{tabular}
\caption{The analysis cuts on a set of kinematic variables at 14 TeV (27 TeV) grouped by the benchmarks of Table~\ref{tab6}. Notice that with the exception of $m_{\ell\ell}$ harder cuts are applied at 27 TeV. Entries with a dash (-) mean that no requirement on the variable is considered. Also shown at the bottom are the required integrated luminosities for discovery at 14 TeV and 27 TeV. Entries with `NV' mean that the point is not visible at the corresponding center-of-mass energy. }
\label{tab9}
\end{table}

\subsection{Results}

\begin{figure}[H]
 \centering
 \includegraphics[width=0.49\textwidth]{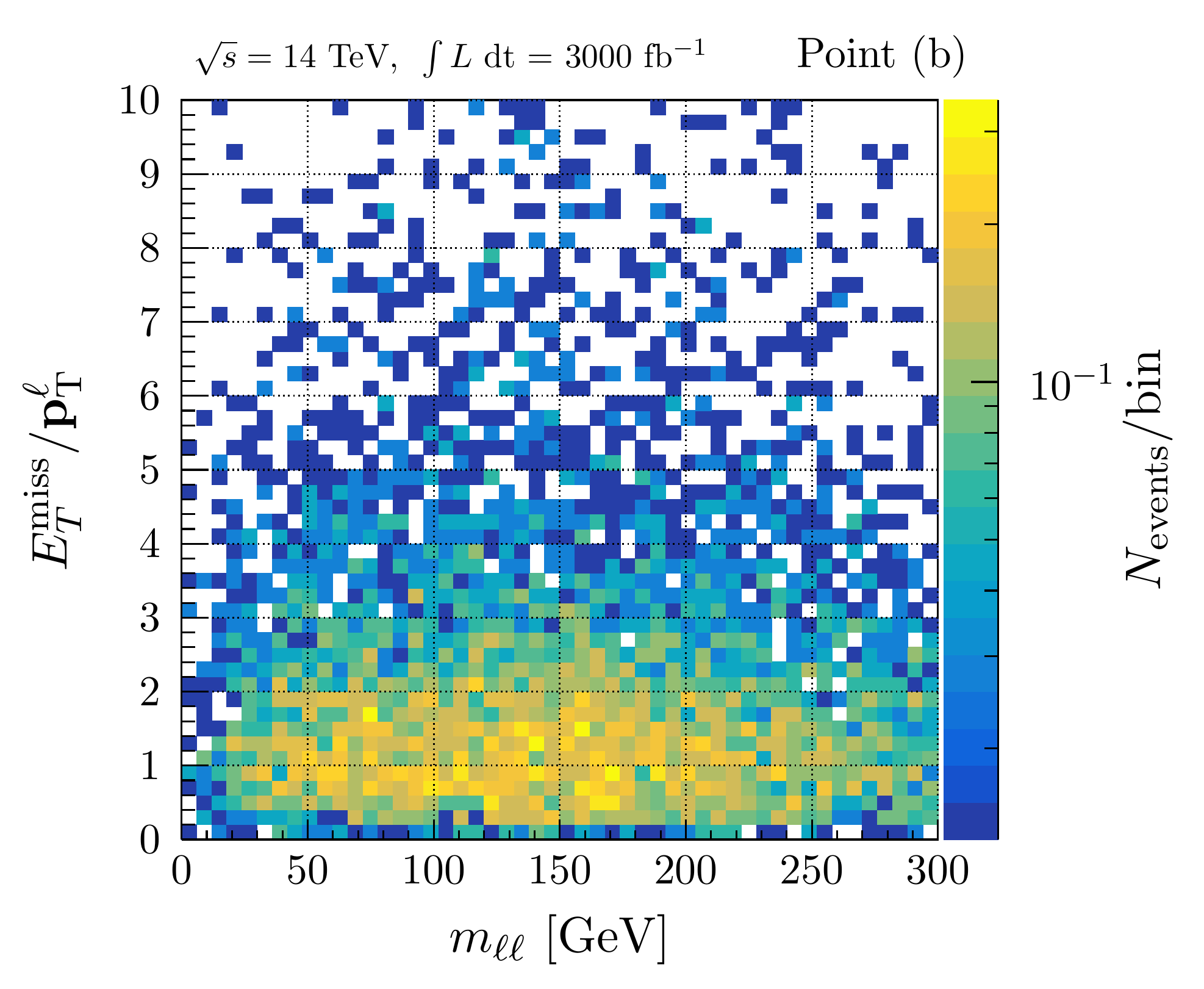}
 \includegraphics[width=0.49\textwidth]{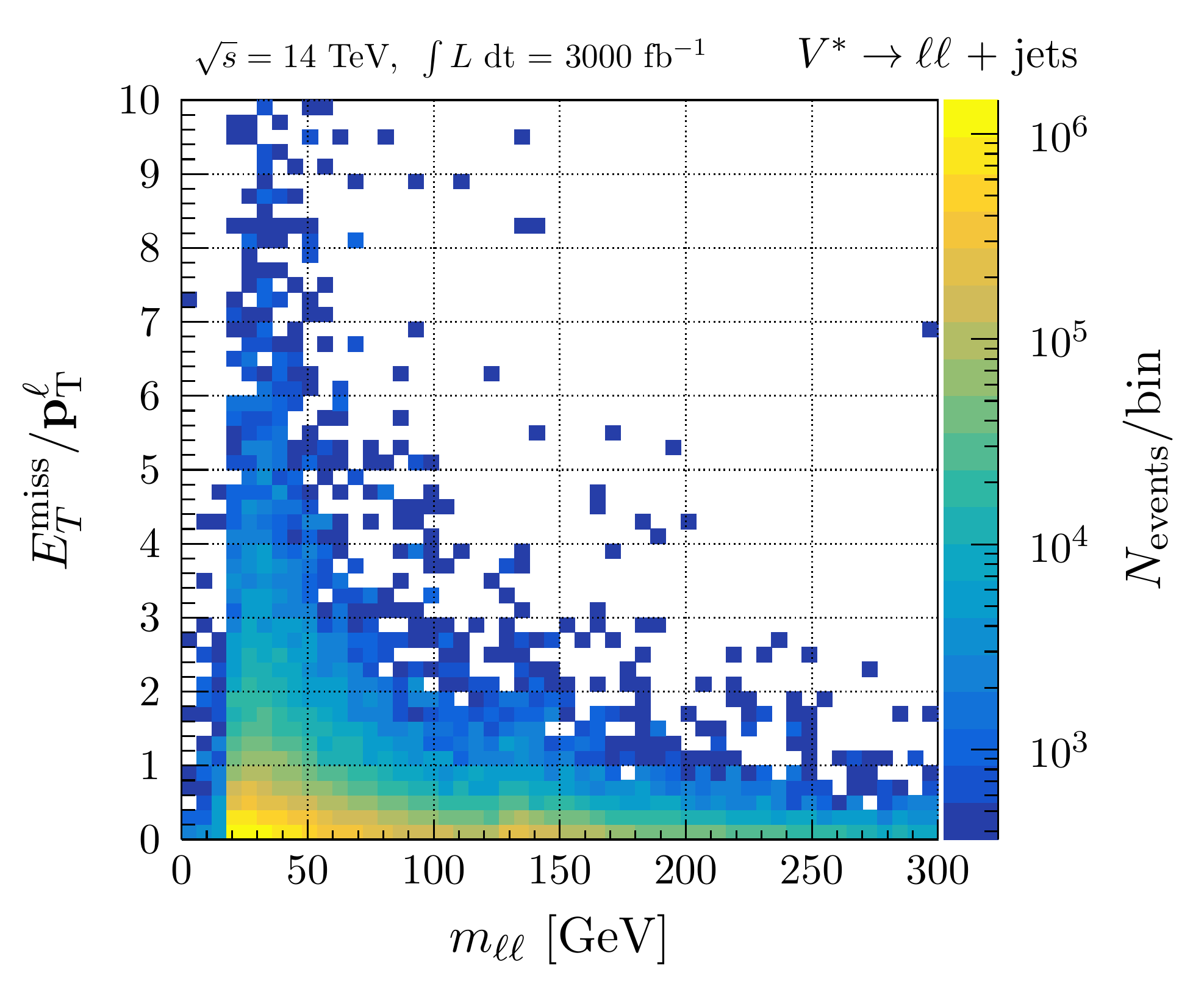} \\
 \includegraphics[width=0.49\textwidth]{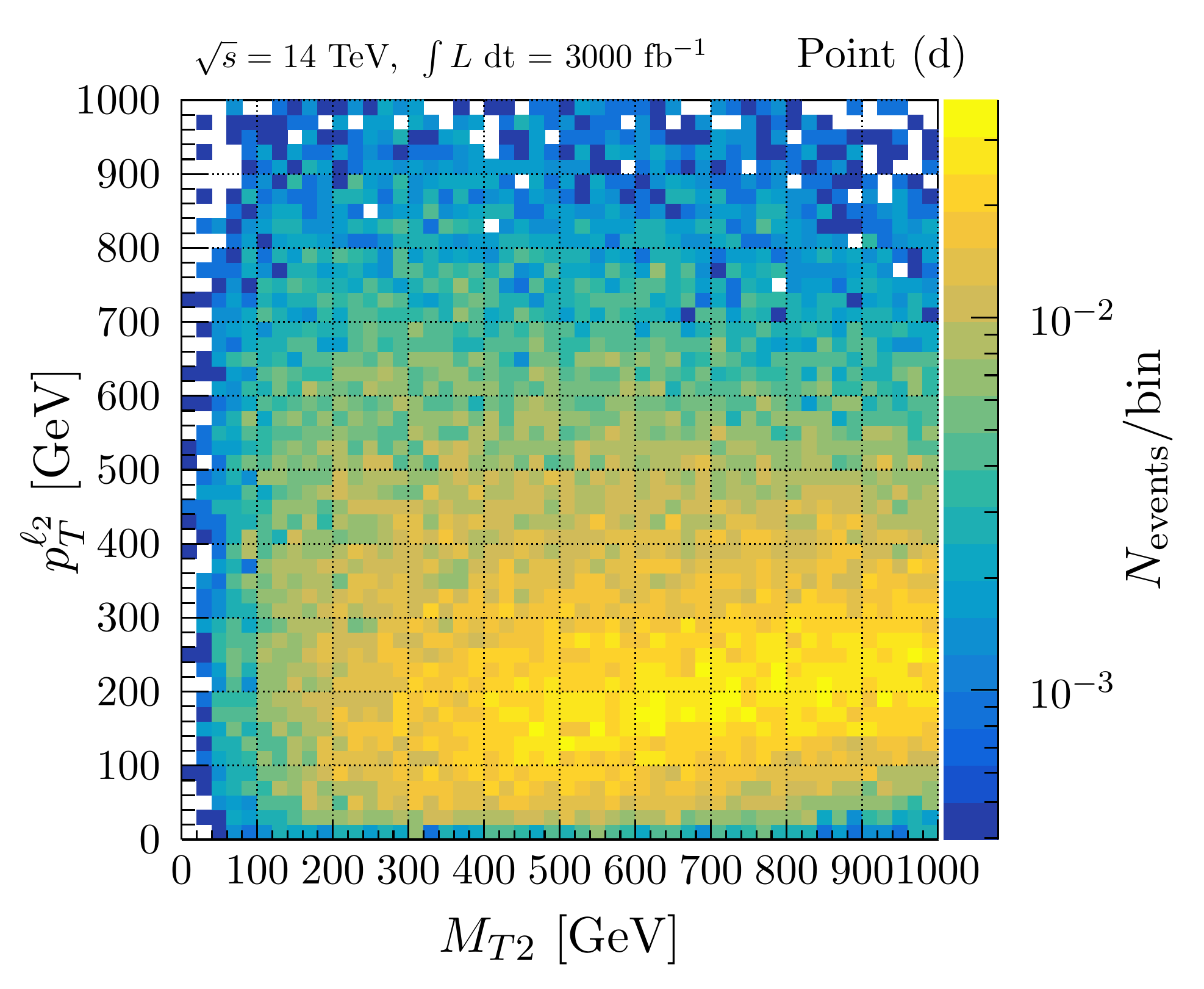}
 \includegraphics[width=0.49\textwidth]{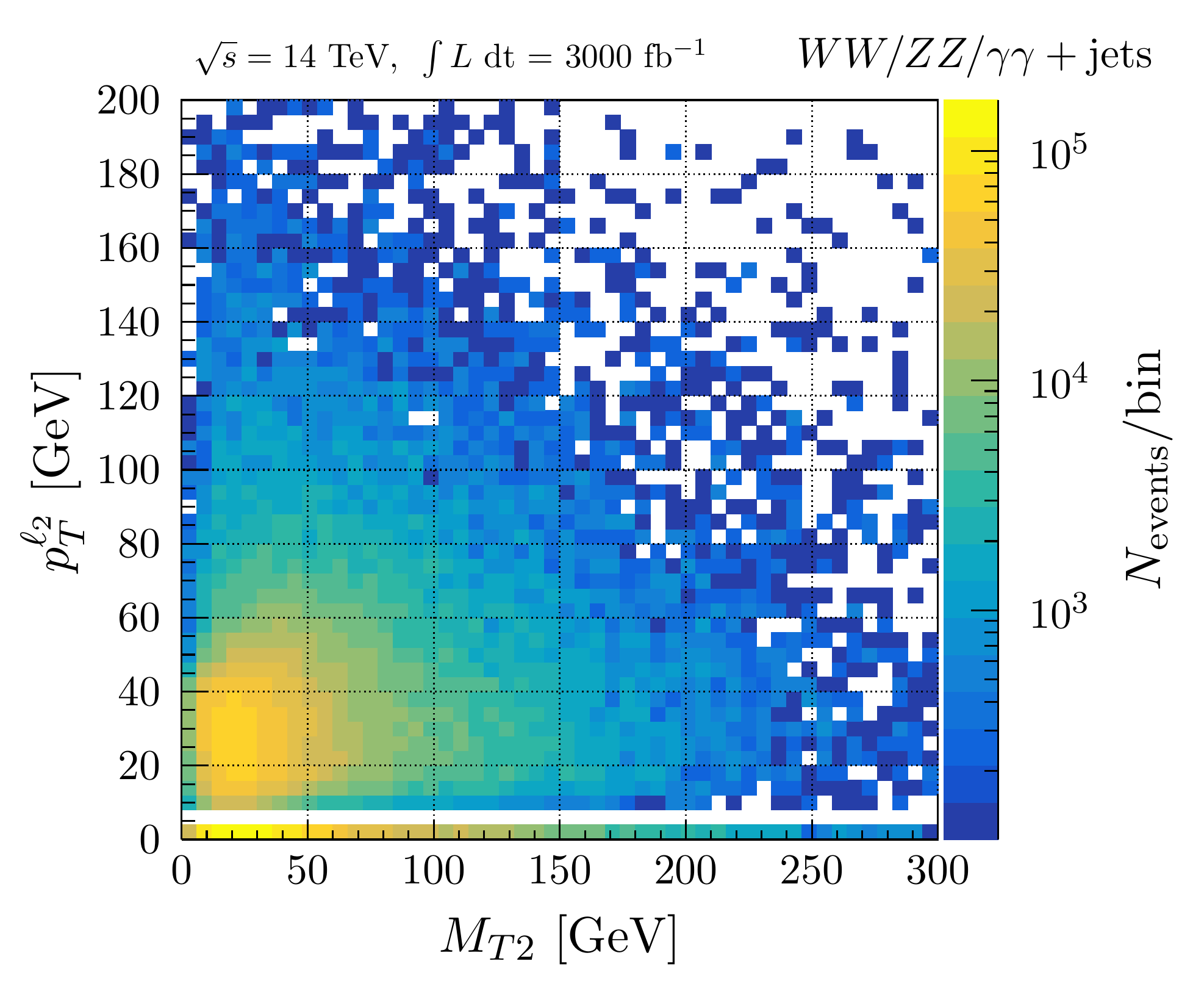} \\
 \includegraphics[width=0.49\textwidth]{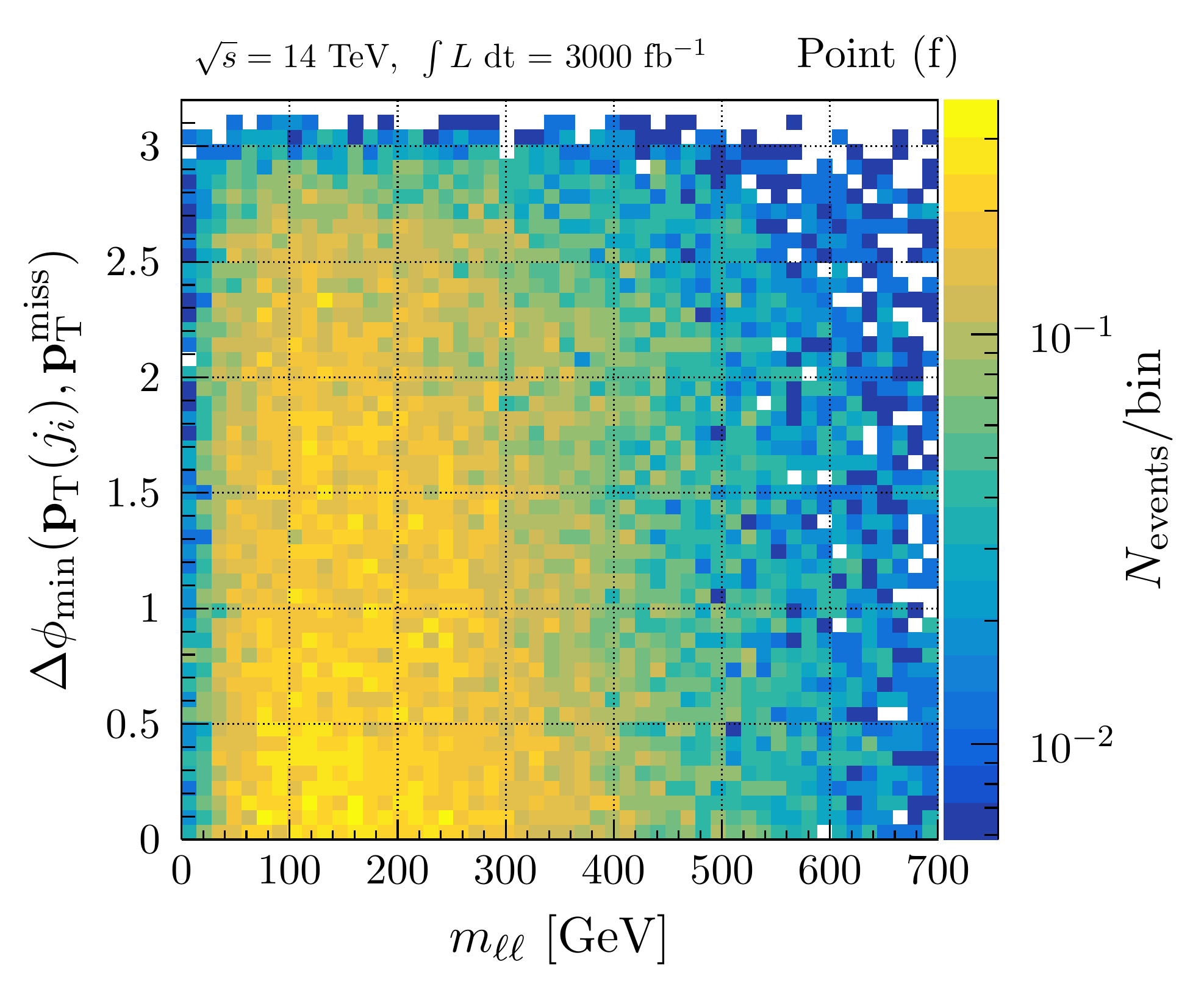}
 \includegraphics[width=0.49\textwidth]{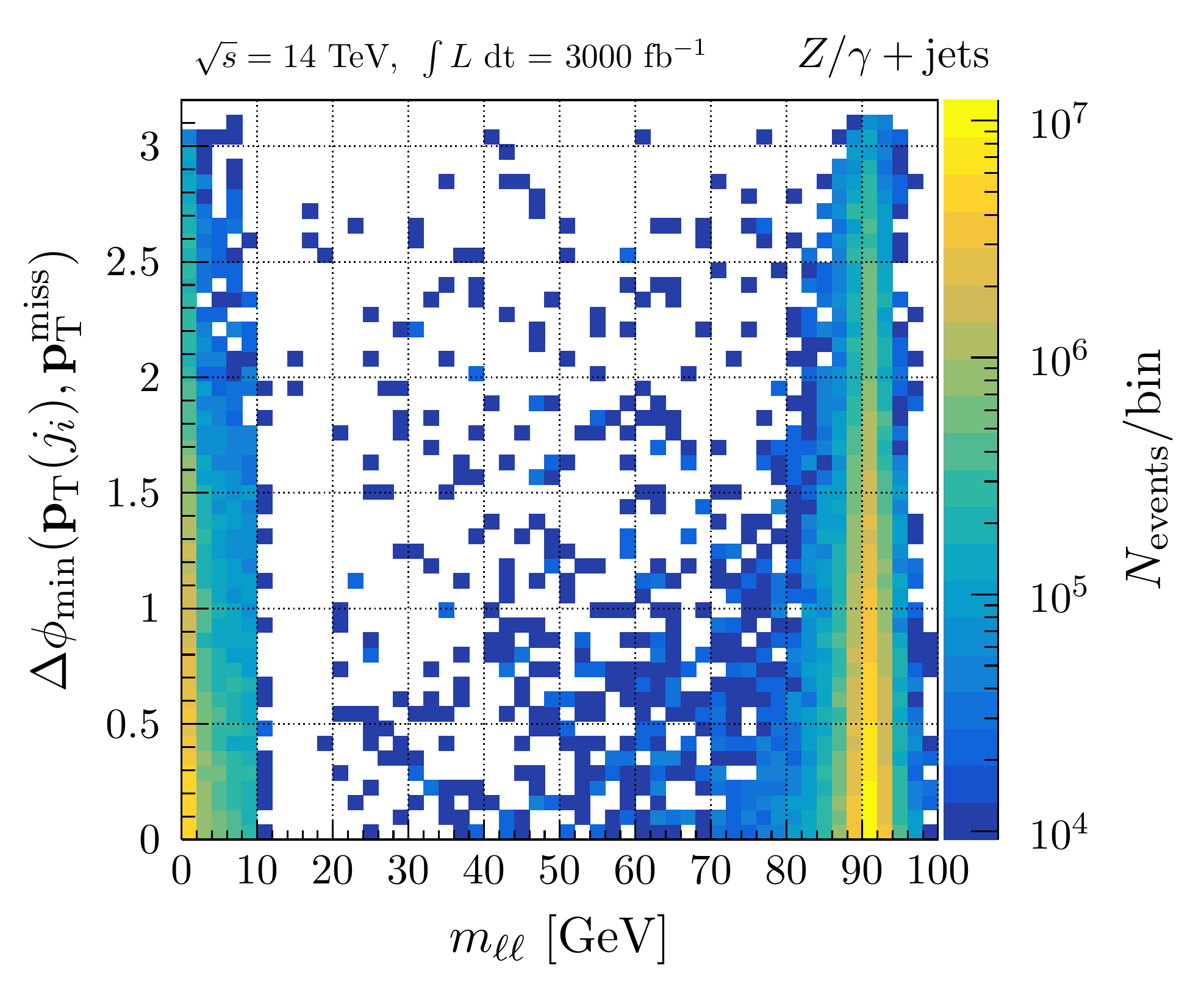}
 \caption{Two dimensional plots in select kinematic variables with the number of events on the color axis. Top panels: $E^{\rm miss}_T/\textbf{p}^{\ell}_{\rm T}$ vs the dilepton invariant mass for benchmark (b) (left) and off-shell vector boson background (right). Middle panels: the subleading lepton transverse momentum vs $M_{T2}$ for benchmark (d) (left) and the diboson background (right). Bottom panels: $\Delta\phi_{\rm min}(\textbf{p}_{\rm T}(j_i),\textbf{p}^{\rm miss}_{\rm T})$ vs the dilepton invariant mass for benchmark (f) (left) and $Z/\gamma+\text{jets}$ background (right). }
	\label{fig4}
\end{figure}

We begin by discussing the benchmarks (d), (g) and (h) which belong to Case 3. Here the mass splitting between the slepton and the neutralino is large, ranging from 300 GeV to 600 GeV, which produces very energetic leptons. For those benchmarks, the sleptons decay to a light lepton and a neutralino with a 100\% branching ratio (see Table~\ref{tab8}) which makes for a clean final state. The most effective kinematic variables for this case are $M_{T2}$ and $p_T^{\ell_2}$ where the latter is the transverse momentum of the subleading lepton. We present two-dimensional plots in these variables in the middle panels of Fig.~\ref{fig4}. The left panel depicts point (d) and the right one is the dominant diboson background. One can clearly see that the largest number of background events (color axis) are concentrated at small $M_{T2}$ and $p_T^{\ell_2}$ while for the signal larger values are highly populated as well due to the energetic final states. A hard cut on $M_{T2}$ and $p_T^{\ell_2}$ as well as the `DNN response' can reject most of the background events.

Next, we discuss benchmarks (b) and (i) which belong to Case 2. Here the branching ratios to a lepton and a neutralino are smaller, at 31\% and the slepton-neutralino mass gaps are at 85 GeV and 140 GeV, respectively. Such a mass gap is not enough to allow harder cuts on $p_T^{\ell_2}$ and that's why it has been omitted in Table~\ref{tab9}. For this reason, we make use of the leading and subleading transverse momenta of the leptons to reconstruct the total momentum of the system, $\textbf{p}^{\ell}_{\rm T}$, to form the new variable $E^{\rm miss}_T/\textbf{p}^{\ell}_{\rm T}$. Two-dimensional plots in the $E^{\rm miss}_T/\textbf{p}^{\ell}_{\rm T}$ and the dilepton invariant mass, $m_{\ell\ell}$, variables are shown in the top panels of Fig.~\ref{fig4}. The left panel shows the distributions for point (b) while the right one is for dilepton production from off-shell vector bosons. For the background, most of the events lie in the region $E^{\rm miss}_T/\textbf{p}^{\ell}_{\rm T}<2$ and $m_{\ell\ell}<100$ GeV which is the reason for the choice of cuts in Table~\ref{tab9}.

Finally, for point (f) which belongs to Case 1, the branching fraction to a lepton and a neutralino is the smallest compared to its decay to a second neutralino and a chargino. The second neutralino and chargino decay predominantly to a stau which in turn decays to a neutralino and a tau. Hence we are faced with a case of tau-enriched final state which can hadronize forming jets. In our selection, we have rejected b-tagged jets but made no special requirements on tau-tagged jets.  For this particular case, jets (tau-tagged or not) can be used to reject the SM background through the variable $\Delta\phi_{\rm min}(\textbf{p}_{\rm T}(j_i),\textbf{p}^{\rm miss}_{\rm T})$ defined above. In the bottom panels of Fig.~\ref{fig4} we show this variable plotted against $m_{\ell\ell}$ for point (f) (left panel) and the $Z/\gamma$+jets background (right panel). Excluding the region formed by $\Delta\phi_{\rm min}(\textbf{p}_{\rm T}(j_i),\textbf{p}^{\rm miss}_{\rm T})<1$ rad and $m_{\ell\ell}<100$ GeV is effective in reducing the SM background.

\begin{figure}[H]
 \centering
 \includegraphics[width=0.49\textwidth]{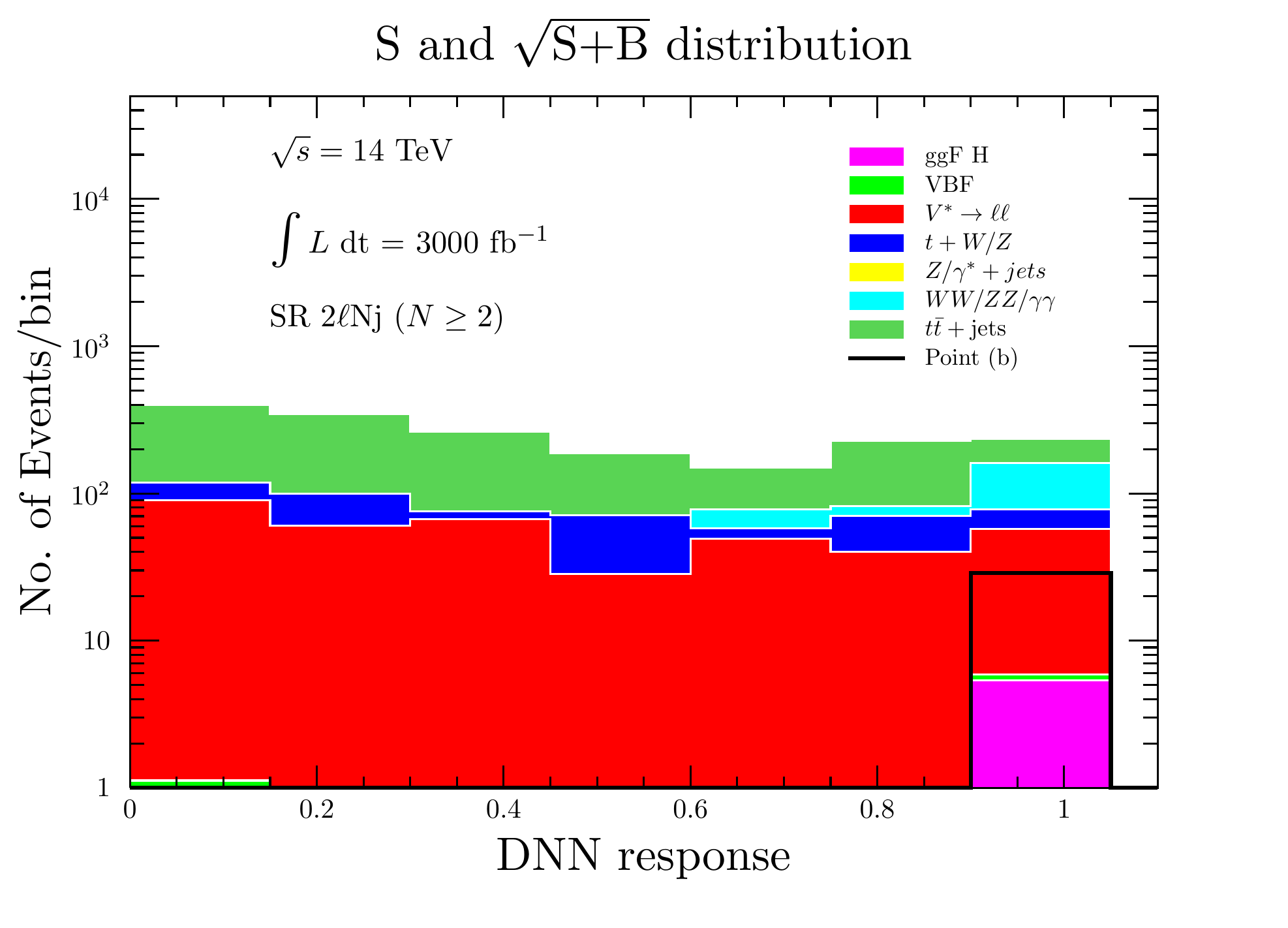}
 \includegraphics[width=0.49\textwidth]{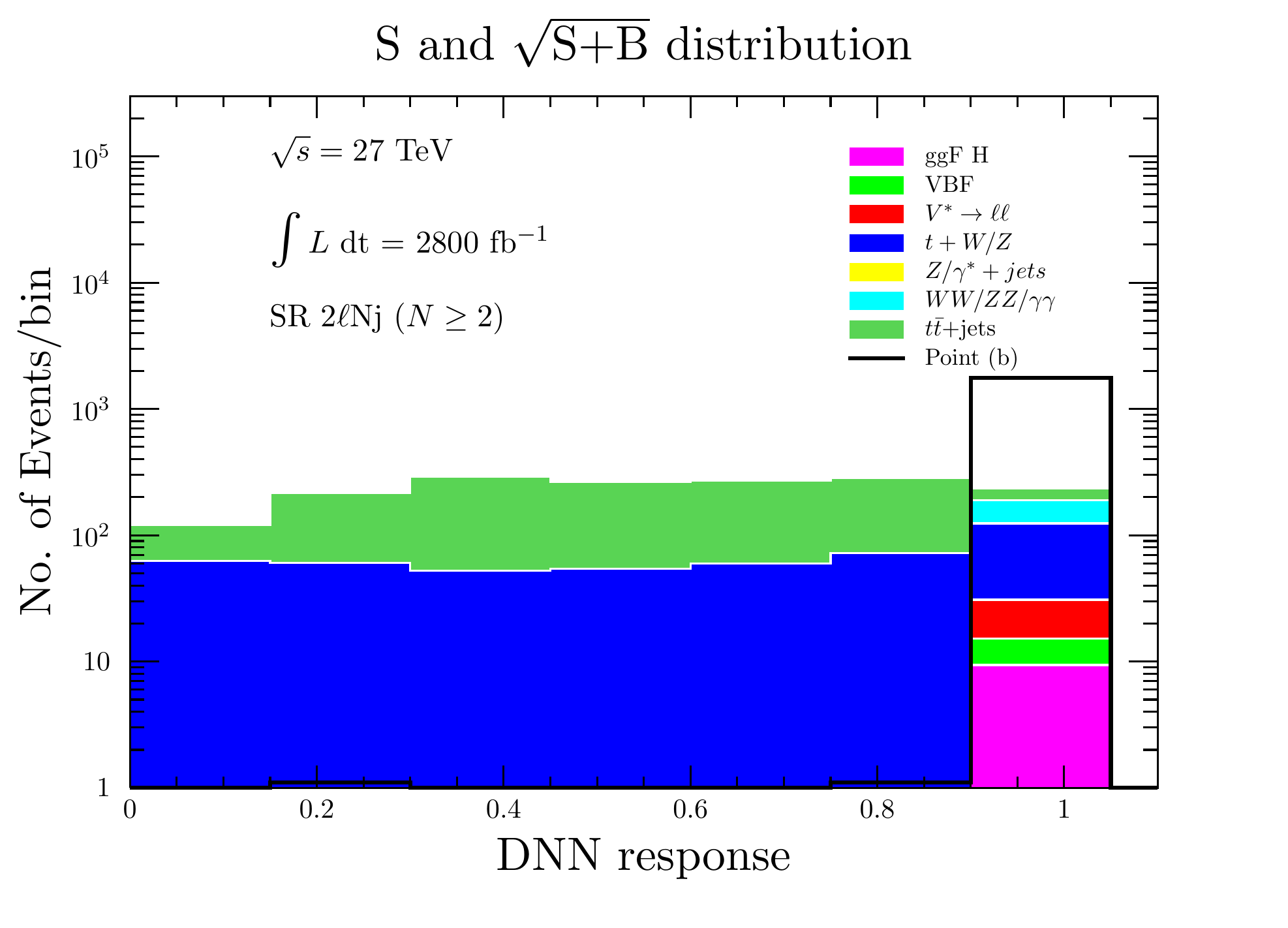} \\
 \includegraphics[width=0.49\textwidth]{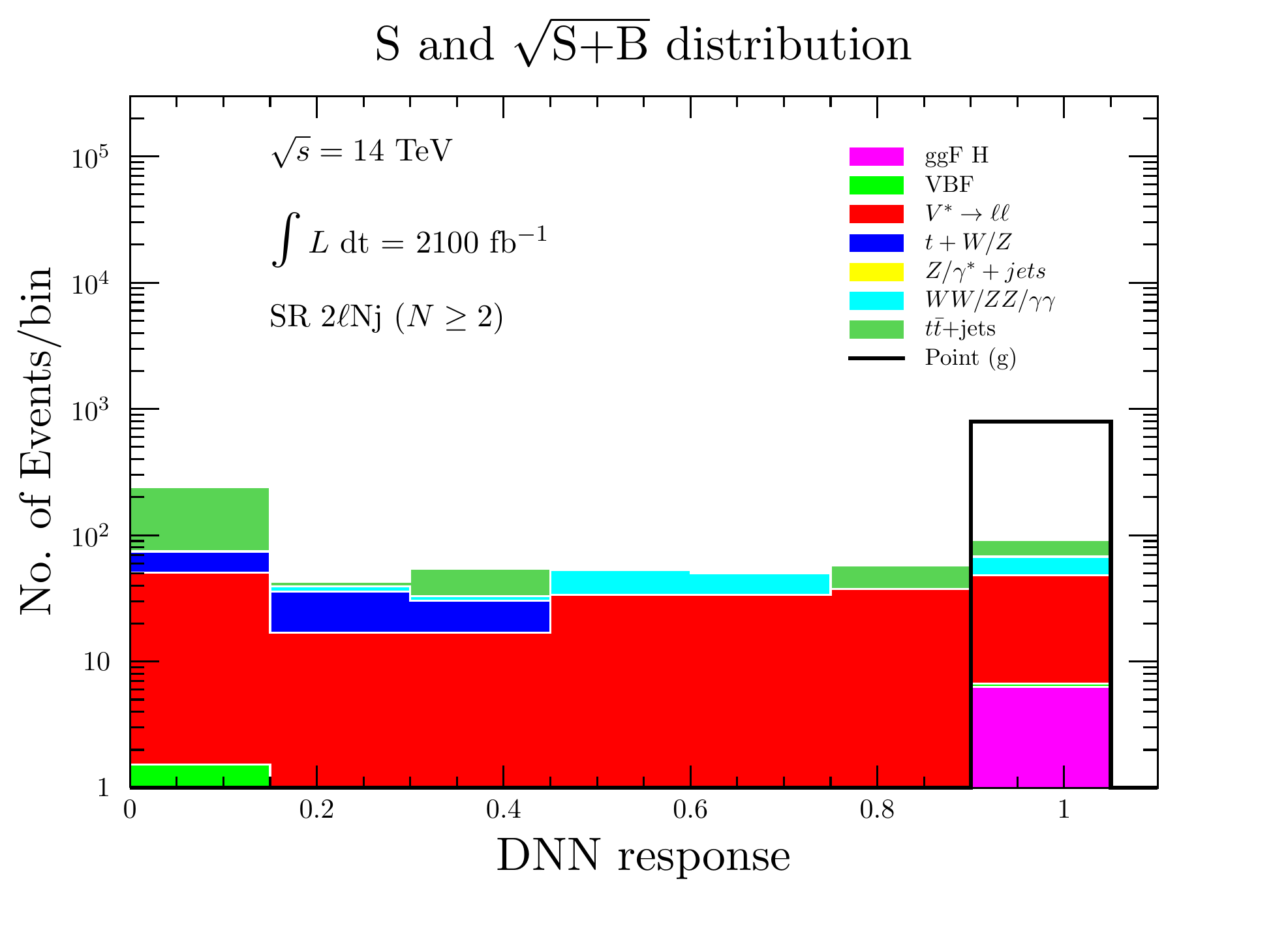}
 \includegraphics[width=0.49\textwidth]{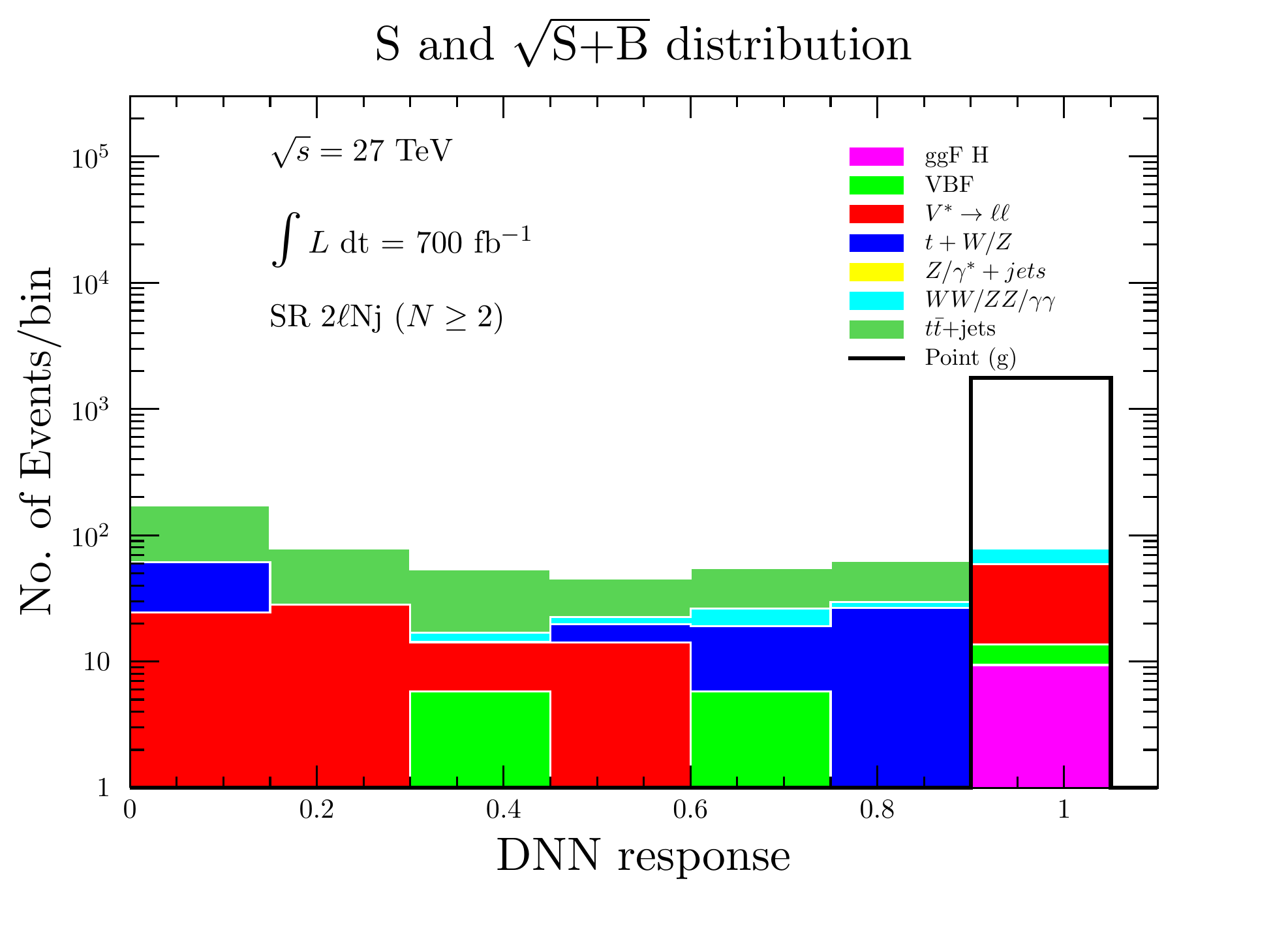}
 \caption{Distributions in the DNN response variable at 14 TeV (left) and 27 TeV (right) for benchmarks (b) (top panels) and (g) (bottom panels). }
 \label{fig5}
\end{figure}

Along with cuts on the variables discussed thus far, the `DNN response' plays an important role. We show in Fig.~\ref{fig5} distributions in this variable after the above cuts have been implemented. The top panel depicts benchmark (b) which shows clearly that at 14 TeV this point cannot be discovered with 3000 fb$^{-1}$ while the signal is in excess over the background near 1 for 2800 fb$^{-1}$ at 27 TeV. The bottom panels show point (g) also at 14 TeV (left) and 27 TeV (right). The benchmark is discoverable at both HL-LHC and HE-LHC but requires smaller integrated luminosity for discovery at HE-LHC (700 fb$^{-1}$) than at HL-LHC (2100 fb$^{-1}$). The evaluated integrated luminosities for discovery at both machines are summarized in the lower part of Table~\ref{tab9}. Entries with `NV' indicate that the benchmark is not discoverable at the corresponding machine. Note that there is a modest improvement in the integrated luminosity at HE-LHC in comparison to HL-LHC but the former is expected to gather data at the rate of $\sim 820$ fb$^{-1}$ per month, so most of those points will be discoverable within the first two to three months of run. Note that points (f), (g), (h) and (i) are discoverable at both machines while (b) and (d) can only be discoverable at HE-LHC.

We note that recently several works have come out regarding a SUSY explanation of the Fermilab muon $g-2$~\cite{Iwamoto:2021aaf,Gu:2021mjd,VanBeekveld:2021tgn,Yin:2021mls,Wang:2021bcx,Cao:2021tuh,Chakraborti:2021dli,Cox:2021gqq,Han:2021ify,Baum:2021qzx,Ahmed:2021htr,Baer:2021aax,Endo:2021zal,Ibe:2021cvf,Chakraborti:2021bmv}.

\section{Conclusion\label{sec5}}
 In this work we have investigated if high scale models can produce Yukawa coupling unification
 consistent with the Fermilab muon $g-2$ result. We used a neural network to investigate the parameter
 space of a class of $\mathsf{SO(10)}$ models where Yukawa couplings arise from the cubic as well
 as the quartic interactions. As in a recent work it is found that the preferred parameter
 space lies in a region where  gluino-driven radiative breaking of the electroweak symmetry occurs. The model produces a split spectrum consisting of a light sector and a heavy sector.
 The light sector contains light sleptons and light weakinos, and the heavy sector contains
 the gluino, the squarks and the heavy Higgs. The masses of the light sparticles lie in the
 few hundred GeV range and are accessible at the LHC. With the help of a deep neural network, we carried out a dedicated search of sleptons in the two-lepton final state at HL-LHC and HE-LHC. It is found that most of the considered benchmarks are discoverable  within the optimal integrated luminosity of HL-LHC while all of them are discoverable at HE-LHC with less integrated luminosities.

\textbf{Acknowledgments:} The research of AA was supported by the BMBF under contract 05H18PMCC1, while the research of PN was supported in part by the NSF Grant PHY-1913328.

\section{Appendix: Contributions to Yukawas from higher dimensional operators}

  In this appendix we give the contributions $\delta h_b, \delta h_t, \delta h_\tau$  to the Yukawas that arise from higher dimensional operators where
  \begin{align}
\delta h_t=\delta  h_t^{(1)} + \delta h_t^{(2)}+\delta h_t^{(3)},~~\delta h_b=\delta  h_b^{(1)} + \delta h_b^{(2)}+ \delta h_b^{(3)} ,~~
\delta h_\tau=\delta  h_\tau^{(1)} + \delta h_\tau^{(2)}+\delta h_\tau^{(3)}\,.
\label{yuk-gut}
\end{align}
Here $\delta  h^{(1)}$ is the contribution arising from $W_4^{(1)}$,
 $\delta  h^{(2)}$ is the contribution arising from $W_4^{(2)}$, and  $\delta  h^{(3)}$ is the contribution arising from $W_4^{(3)}$. The explicit forms of these are given below~\cite{Aboubrahim:2020dqw}.

Thus $W_4^{(1)}$ gives the following contribution to the third generation Yukawas
\begin{eqnarray}
\delta h^{(1)}_{t}&=& \frac{if^{(1)}}{60\sqrt{2}M_{c}}\left(\sum_{r=1}^2b_rU_{d_{r1}}\right)\left[\frac{5\sqrt{3}}{2}\mathcal V_{75_{_{{210}}}}-4\sqrt{15}\mathcal V_{24_{_{{210}}}}-8\sqrt{15}\mathcal V_{1_{_{{210}}}}\right], \label{q&l masses from quartic coupling 1a} \\
\delta h^{(1)}_{b}&=& \frac{if^{(1)}}{60\sqrt{2}M_{c}}\left(\sum_{r=1}^2b_rV_{d_{r1}}\right)\left[\frac{\sqrt{20}}{3}\mathcal V_{75_{_{{210}}}}-20\sqrt{\frac{5}{3}}\mathcal V_{24_{_{{210}}}}\right],   \label{q&l masses from quartic coupling 1b}\\
\delta h^{(1)}_{\tau}&=& \frac{if^{(1)}}{60\sqrt{2}M_{c}}\left(\sum_{r=1}^2b_rV_{d_{r1}}\right)\left[20\sqrt{3}\mathcal V_{75_{_{{210}}}}-20\sqrt{15}\mathcal V_{24_{_{{210}}}}\right],\label{q&l masses from quartic coupling 1c}
\end{eqnarray}

 The contribution of  $W_4^{(2)}$ to the third generation Yukawas is given by
\begin{align}
\delta h^{(2)}_{t}=&-\frac{i f^{(2)}}{120M_{c}} \left[\frac{10}{3}\sqrt{\frac{2}{3}}\mathcal V_{75_{_{{210}}}}U_{d_{61}}
+\frac{5}{3}\sqrt{\frac{10}{3}}\mathcal V_{24_{_{{210}}}}U_{d_{61}}+6\sqrt{5}\mathcal V_{24_{_{{210}}}}U_{d_{31}}-8\sqrt{5}\mathcal V_{1_{_{{210}}}}U_{d_{31}}\right],\label{q&l masses from quartic coupling 2a} \\
\delta h^{(2)}_{b}=&-\frac{i f^{(2)}}{120M_{c}} \Bigg[-\frac{20}{3}\sqrt{\frac{2}{3}}\mathcal V_{75_{_{{210}}}}V_{d_{61}}-\frac{20}{3}\mathcal V_{75_{_{{210}}}}V_{d_{31}}
-\frac{1}{3}\sqrt{\frac{10}{3}}\mathcal V_{24_{_{{210}}}}V_{d_{61}}-\frac{10\sqrt{5}}{3}\mathcal V_{24_{_{{210}}}}V_{d_{31}} \nonumber \\
&\hspace{2cm}-4\sqrt{\frac{10}{3}}\mathcal V_{1_{_{{210}}}}V_{d_{61}}\Bigg],\label{q&l masses from quartic coupling 2b} \\
\delta h^{(2)}_{\tau}=& -\frac{i f^{(2)}}{120M_{c}} \Bigg[-20\sqrt{\frac{2}{3}}\mathcal V_{75_{_{{210}}}}V_{d_{61}}-20\mathcal V_{75_{_{{210}}}}V_{d_{31}}
-\sqrt{\frac{10}{3}}\mathcal V_{24_{_{{210}}}}V_{d_{61}}-10\sqrt{5}\mathcal V_{24_{_{{210}}}}V_{d_{31}} \nonumber \\
&\hspace{2cm}-4\sqrt{30}\mathcal V_{1_{_{{210}}}}V_{d_{61}}\Bigg].\label{q&l masses from quartic coupling 2c}
\end{align}
Finally, the contribution of  $W_4^{(3)}$ to the third generation Yukawas is given by
\begin{align}
\delta h_t^{(3)}&=-\frac{3i}{8}\frac{f^{(3)}}{M_c}\left[\frac{2}{3}\sqrt{\frac{2}{3}}\mathcal V_{75_{210}}U_{d_{61}}
+\frac{1}{3}\sqrt{\frac{10}{3}}\mathcal V_{24_{210}}U_{d_{61}}-\frac{2}{\sqrt{5}}\mathcal V_{24_{210}}U_{d_{31}}+\frac{8}{3\sqrt{5}}\mathcal V_{1_{210}}U_{d_{31}}\right],~~~\label{q&l masses from quartic coupling 3a}\\
\delta h_b^{(3)}&=-\frac{3i}{8}\frac{f^{(3)}}{M_c}\left[\frac{2}{3}\sqrt{\frac{2}{3}}\mathcal V_{75_{210}}V_{d_{61}}
+\frac{1}{3}\sqrt{\frac{10}{3}}\mathcal V_{24_{210}}V_{d_{61}}-\frac{2}{\sqrt{5}}\mathcal V_{24_{210}}V_{d_{31}}+\frac{8}{3\sqrt{5}}\mathcal V_{1_{210}}V_{d_{31}}\right],\label{q&l masses from quartic coupling 3b}\\
\delta h_{\tau}^{(3)}&=\frac{3i}{8}\frac{f^{(3)}}{M_c}\left[\frac{2}{3}\sqrt{\frac{2}{3}}\mathcal V_{75_{210}}V_{d_{61}}
+\frac{1}{3}\sqrt{\frac{10}{3}}\mathcal V_{24_{210}}V_{d_{61}}-\frac{2}{\sqrt{5}}\mathcal V_{24_{210}}V_{d_{31}}+\frac{8}{3\sqrt{5}}\mathcal V_{1_{210}}V_{d_{31}}\right].\label{q&l masses from quartic coupling 3c}
\end{align}
The total  Yukawas are the sum of the contributions from the cubic and from the quartic terms at the GUT scale as given in Eq.~(\ref{yuksum}).

\end{document}